\documentclass[titlepage]{csetr}

\usepackage{graphicx}

\usepackage[obeyspaces]{url}

\usepackage{fancyvrb}

\usepackage{geometry}
\usepackage{float}
\usepackage{setspace,supertabular}
\usepackage{longtable}
\usepackage[usenames]{color}
\usepackage{verbatim}
\usepackage{subfigure}
\usepackage{amsmath}
\usepackage[table]{xcolor}

\usepackage{lscape}
\usepackage{changepage}
\usepackage{tabularx}

\usepackage{
tikz,
relsize,
amsmath,
booktabs,
tikz
}

\usepackage[
labelfont=sf,
hypcap=false,
format=hang,
width=0.8\columnwidth
]{caption}

\usepackage[T1]{fontenc}
\usepackage[utf8]{inputenc}
\usepackage{lmodern}

\renewcommand{\and}{\hspace{.5cm}}

\trnumber{201617}
\trdate{December 2016}

\title{%
  Data Curation APIs
}
\author{%
  Seyed-Mehdi-Reza Beheshti$^1$ \and %
  Alireza Tabebordbar$^1$ \and \\%
  Boualem Benatallah$^1$ \and %
  Reza Nouri$^1$\\[2em]
  $^1\, $University of New South Wales, Australia \\%
  \email{\{sbeheshti,a.tabebordbar,boualem,snouri\}@cse.unsw.edu.au}\\[3cm]
}

\date{}
\begin{document}
\maketitle

\begin{abstract}

Understanding and analyzing big data is firmly recognized as a powerful and strategic priority.
For deeper interpretation of and better intelligence with big data, it is important to transform raw data (unstructured, semi-structured and structured data sources, e.g., text, video, image data sets) into curated data: contextualized data and knowledge that is maintained and made available for use by end-users and applications.
In particular, data curation acts as the glue between raw data and analytics, providing an abstraction layer that relieves users from time consuming, tedious and error prone curation tasks. In this context, the \emph{data curation} process becomes a vital analytics asset for increasing added value and insights.

In this paper, we identify and implement a set of curation APIs and make them available\footnote{We encourage researchers/developers to cite our work if you have used our APIs, libraries, tools or datasets.} (on GitHub\footnote{https://github.com/unsw-cse-soc/Data-curation-API.git}) to researchers and developers to assist them transforming their raw data into curated data.
The curation APIs enable developers to easily add features - such as
extracting keyword, part of speech, and named entities such as Persons, Locations, Organizations, Companies, Products, Diseases, Drugs, etc.;
providing synonyms and stems for extracted information items leveraging lexical knowledge bases for the English language such as WordNet;
linking extracted entities to external knowledge bases such as Google Knowledge Graph and Wikidata;
discovering similarity among the extracted information items, such as calculating similarity between string, number, date and time data;
classifying, sorting and categorizing data into various types, forms or any other distinct class;
and indexing structured and unstructured data -
into their applications.

\end{abstract}

\section{Introduction}

Understanding and analyzing big data is firmly recognized as a powerful and strategic priority.
For deeper interpretation of and better intelligence with big data, it is important to transform raw data (unstructured, semi-structured and structured data sources, e.g., text, video, image data sets) into curated data: contextualized data and knowledge that is maintained and made available for use by end-users and applications~\cite{CDCR1,CDCR2}.
In particular, data curation acts as the glue between raw data and analytics, providing an abstraction layer that relieves users from time consuming, tedious and error prone curation tasks.

Data curation involves identifying relevant data sources, extracting data and knowledge, cleaning, maintaining, merging, enriching and linking data and knowledge.
For example, consider a tweet in the Twitter\footnote{https://twitter.com/}~\cite{kwak2010twitter}: a microblogging service that enable users tweet about any topic within the 140-character limit and follow others to receive their tweets. It is possible to extract various information from a single tweet text such as keywords, part of speech, named entities, synonyms and stems\footnote{https://en.wikipedia.org/wiki/Word\_stem}. Then it is possible to link the extracted information to external knowledge graphs to enrich and annotate the raw data~\cite{Galaxy,socialRating}.
Figure~\ref{fig:twitterfold} illustrates possible information items that can be extracted from a tweet. Later, these information can be used to provide deeper interpretation of and better intelligence with the huge number of tweets in Twitter: every second, on average, around 6,000 tweets are tweeted on Twitter, which corresponds to over 350,000 tweets sent per minute, 500 million tweets per day and around 200 billion tweets per year.

Considering this example, the main challenge would be to understand the extremely large (potentially linked) open and private datasets.
In this context, the data curation will enable data scientists to understand and analyze the big data, understand it better and extract more value and insights.
In particular, applying analytics - such as process analytics~\cite{ProcessAnalytics,DAPD,Zakaria,BusinessProcessParadigms,rocessMatchingTechniques,ProcessQuery,BPM,ICSOCJohn},
information networks analysis~\cite{GraphAnalytics,GraphSurvey,Dream,TPCTC,FPSPARQL,BPSPARQL,Bilal2,Bilal1},
data and metadata analysis~\cite{Rajabi2016,crossCuttingAspects,Provenance,resourceDiscovery,ProcessDataAnalysis} -
to the curated data can reveal patterns, trends, and associations and consequently increase the added value and insights of the raw data.
Trending applications include but not limited to
improve government services~\cite{criado2013government,chen2012business};
predict intelligence activities~\cite{van2014datafication,fader2014open};
unravel human trafficking activities~\cite{burke2013human,crowdsourcingSystems,PeopleEvaluation};
understand impact of news on stock markets~\cite{boudoukh2013news,OnlineRatingSystems1};
analysis of financial risks~\cite{carmona2014statistical,OnlineRatingSystems2};
accelerate scientific discovery~\cite{towns2014xsede,heer2015predictive};
as well as to improve national security and public health~\cite{tene2012big,kamel2016instagram}.

In this paper, we identify and implement a set of curation APIs and make them available (on GitHub\footnote{https://github.com/unsw-cse-soc/Data-curation-API.git}) to researchers and developers to assist them transforming their raw data into curated data.
The curation APIs enable developers to easily add features - such as
extracting keyword, part of speech, and named entities such as Persons, Locations, Organizations, Companies, Products, Diseases, Drugs, etc.;
providing synonyms and stems for extracted information items leveraging lexical knowledge bases for the English language such as WordNet;
linking extracted entities to external knowledge bases such as Google Knowledge Graph\footnote{https://developers.google.com/knowledge-graph/} and Wikidata\footnote{https://www.wikidata.org/};
discovering similarity among the extracted information items, such as calculating similarity between string, number, date and time data;
classifying, sorting and categorizing data into various types, forms or any other distinct class;
and indexing structured and unstructured data -
into their applications.

The rest of the paper is organized as follows.
In Section~\ref{CurationServices}, we introduce the Curation APIs that have been implemented to assist the researchers and developers easily add features into their applications and to facilitate the task of transforming their raw data into curated data.
In Section~\ref{MotivatingScenario} we illustrate a motivating scenario to curate Twitter data: contextualized Twitter data and knowledge that is maintained and made available for use by end-users and applications.
Finally, in the Appendix, we present the specification, implementation and evaluation of the curation APIs, made available on GitHub.
We encourage researchers and developers to cite our work if you have used our APIs, libraries, tools or datasets.

\section{The Need for Curation Services}
\label{CurationServices}

The successful transformation of data into knowledge requires going beyond mere extraction and traditional database-like manipulation operators such as filtering, projection and join. Accordingly, to enhance the curation process, we propose a framework for data curation feature engineering: this refers to characterizing variables that grasp and encode information from raw or curated data, thereby enabling to derive meaningful inferences from data. An example of a feature is `mentions of a person in data items like tweets, news articles and social media posts'. Identifying features and algorithms to realize them is a critical yet tedious task: they must be chosen wisely and implemented accurately. For that reason, the engineering process of features requires an evolving and iterative process consisting of: selection; computation and performance evaluation.
We propose that features will be implemented and available as uniformly accessible data curation Micro-Services: functions or pipelines implementing features. In particular, we will support a concrete set of features~\cite{anderson2013brainwash,seide2011feature}, organized in categories such as: Extracting, Classifying, Linking, and Enriching algorithms. For example, a classifier algorithm~\cite{duda2012pattern,aggarwal2012survey,aggarwal2014data,thornton2013auto} (e.g. SVM or kNN) could be employed to classify tweets into various subject-matter categories. In particular, we implement a set of extraction and analysis services to assist analysts in being able to apply further operations to raw data and gain more insight from the data~\cite{Galaxy}. In the following, we explain a set of micro-services that we have implemented.

\begin{figure} [t]
\centering
\includegraphics[width=1.0\textwidth]{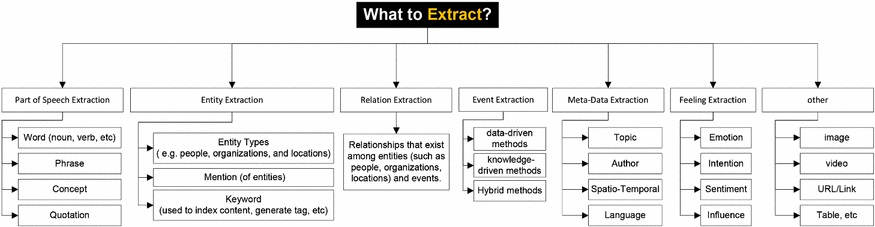}
\caption{A taxonomy of the information that can be extracted from unstructured/semi-structured documents~\cite{CDCR1}.}
\label{fig:extraction}
\end{figure}

\subsection{Extraction Services}

The majority of the digital information produced globally is present in the form of web pages, text documents, news articles, emails, and presentations expressed in natural language text. Collectively, such data is termed unstructured as opposed to structured data that is normalized and stored in a database. The domain of Information Extraction (IE) is concerned with identifying information in unstructured documents and using it to populate fields and records in a database. In most cases, this activity concerns processing human language texts by means of Natural Language Processing (NLP)~\cite{CDCR1,sidorov2014syntactic,manning2014stanford}.

\begin{landscape}

\begin{figure}
\centering
\includegraphics[width=1.5\textwidth]{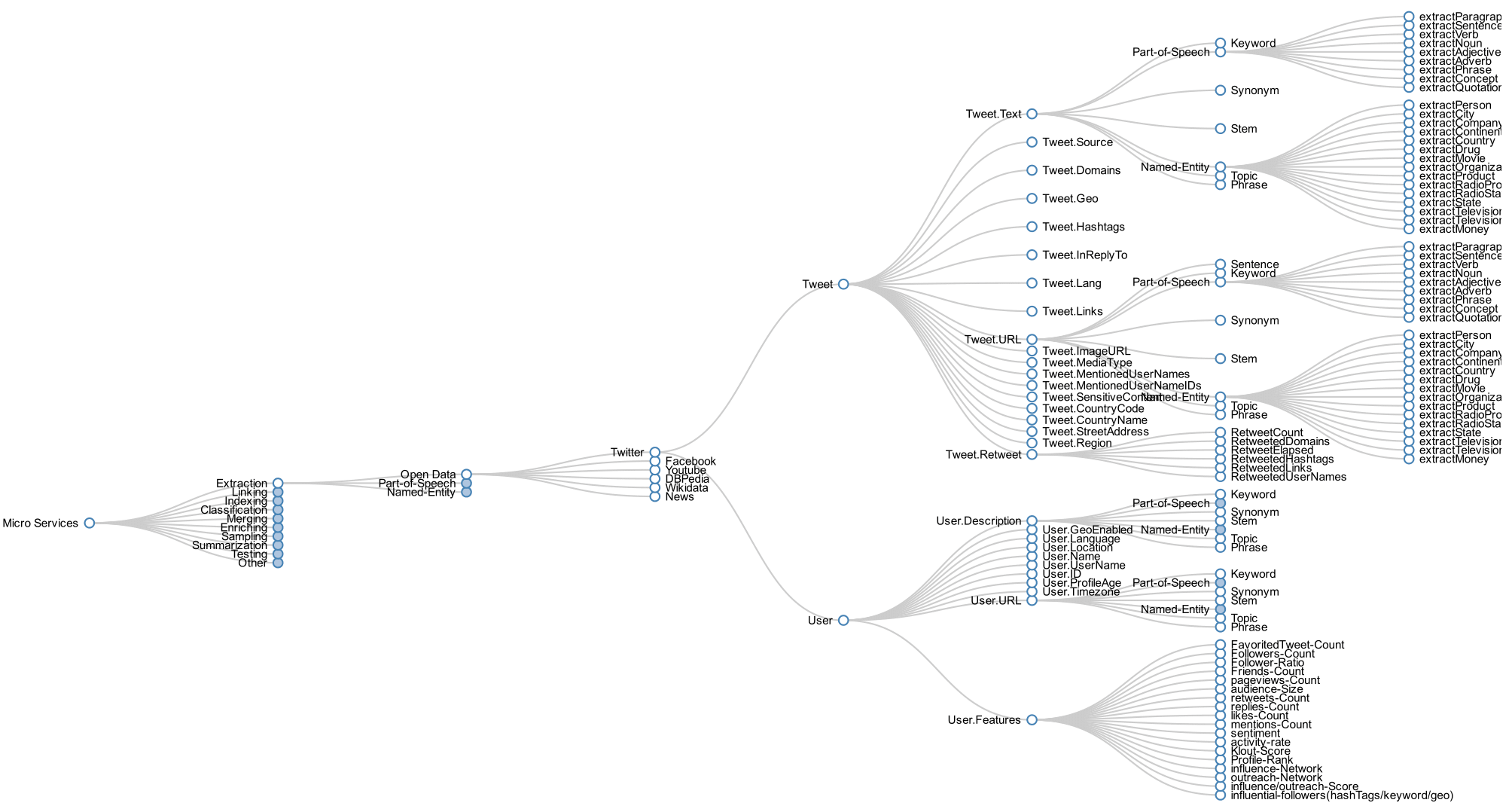}
\caption{Possible information items that can be extracted from a tweet.}
\label{fig:twitterfold}
\end{figure}

\end{landscape}

Figure~\ref{fig:extraction} illustrates a taxonomy of the information that can be extracted from unstructured/semi-structured documents.

\subsubsection{Named Entity}

Named Entity Recognition (NER), also known as Entity Extraction (EE), techniques can be used to locate and classify atomic elements in text into predefined categories such as the names of persons, organizations, locations, expressions of times, quantities, monetary values, and percentages~\cite{CDCR1,nothman2013learning,li2012twiner}. NER is a key part of information extraction systems that supports robust handling of proper names essential for many applications, enables pre-processing for different classification levels, and facilitates information filtering and linking. However, performing coreference, or entity linking, as well as creating templates is not part of an NER task. A basic entity identification task can be defined as follows:
Let ${t_1, t_2, t_3, …, t_n}$ be a sequence of entity types denoted by T and let ${w_1, w_2, w_3, …, w_n}$ be a sequence of words denoted by W, then the identification task can be defined as 'given some W, find the best T'.

In particular, entity identification consists of three subtasks: entity names, temporal expressions, and number expressions, where the expressions to be annotated are 'unique identifiers' of entities (organizations, persons, locations), times (dates, times), and quantities (monetary values, percentages). Most research on entity extraction systems has been structured as taking an unannotated block of text (e.g., ``Obama was born on August 4, 1961, at Gynecological Hospital in Honolulu'') and producing an annotated block of text, such as the following:
\\\\
\scriptsize
<ENAMEX TYPE="PERSON">Obama</ENAMEX> was born on \\
<TIMEX TYPE="DATE">August 4, 1961,</TIMEX> at \\
<ENAMEX TYPE="ORGANIZATION">Gynecological Hospital</ENAMEX> in \\
<ENAMEX TYPE="CITY">Honolulu</ENAMEX>. \\

\normalsize

;where, entity types such as person, organization, and city are recognized.
However, NER is not just matching text strings with pre-defined lists of names. It should recognize entities not only in contexts where category definitions are intuitively quite clear, but also in contexts where there are many grey areas caused by metonymy~\cite{lodge2015modes}. Metonymy is a figure of speech used in rhetoric in which a thing or concept is not called by its own name, but by the name of something intimately associated with that thing or concept. Metonyms can be either real or fictional concepts representing other concepts real or fictional, but they must serve as an effective and widely understood second name for what they represent.
For example, (i) Person vs. Artefact: "The Ham Sandwich (a person) wants his bill." vs "Bring me a ham sandwich."; (ii) Organization vs. Location: "England won the World Cup" vs. "The World Cup took place in England"; (iii) Company vs. Artefact: "shares in MTV" vs. "watching MTV"; and (iv) Location vs. Organization: "she met him at Heathrow" vs. "the Heathrow authorities".

To address these challenges, the Message Understanding Conferences (MUC\footnote{https://en.wikipedia.org/wiki/Message\_Understanding\_Conference}) were initiated and financed by DARPA (Defense Advanced Research Projects Agency) to encourage the development of new and better methods of information extraction. The tasks grew from producing a database of events found in newswire articles from one source to production of multiple databases of increasingly complex information extracted from multiple sources of news in multiple languages. The databases now include named entities, multilingual named entities, attributes of those entities, facts about relationships between entities, and events in which the entities participated. MUC essentially adopted a simplistic approach of disregarding metonymous uses of words, e.g. 'England' was always identified as a location. However, this is not always useful for practical applications of NER, such as in the domain of sports.

MUC defined basic problems in NER as follows: (i) Variation of named entities: for example John Smith, Mr Smith, and John may refer to the same entity; (ii) Ambiguity of named entities types: for example John Smith (company vs. person), May (person vs. month), Washington (person vs. location), and 1945 (date vs. time); (iii) Ambiguity with common words: for example 'may'; and (iv) Issues of style, structure, domain, genre etc. as well as punctuation, spelling, spacing, and formatting. To address these challenges, some of approaches amongst the state-of-the-art in entity extraction proposed four primary steps: Format Analysis, Tokeniser, Gazetteer, Grammar. Figure 3.1 illustrates a simplified process for the NER task. In the following, we provide a brief description of these steps:

\textbf{Format Analysis.} Many document formats contain formatting information in addition to textual content. For example, HTML documents contain HTML tags specifying formatting information such as new line starts, bold emphasis, and font size or style. The first step, format analysis, is the identification and handling of the formatting content embedded within documents that controls the way the document is rendered on a computer screen or interpreted by a software program. Format analysis is also referred to as structure analysis, format parsing, tag stripping, format stripping, text normalization, text cleaning, and text preparation.

\textbf{Tokeniser. }Tokenization is the process of breaking a stream of text up into words, phrases, symbols, or other meaningful elements called tokens. This module is responsible for segmenting text into tokens, e.g., words, numbers, and punctuation. The list of tokens becomes input for further processing such as parsing or text mining.

\textbf{Gazetteer.} This module is responsible for categorizing the type and scope of the information presented. In particular, a gazetteer is a geographical dictionary or directory, an important reference for information about places and place names. It typically contains information concerning the geographical makeup of a country, region, or continent as well as the social statistics and physical features, such as mountains, waterways, or roads. As an output, this module will generate a set of named entities (e.g., towns, names, and countries) and keywords (e.g., company designators and titles).

\textbf{Grammar.} This module is responsible for hand-coded rules for named entity recognition. NER systems are able to use linguistic grammar-based techniques as well as statistical models. Handcrafted grammar-based systems typically obtain better precision, but at the cost of lower recall and months of work by experienced computational linguists. Statistical NER systems typically require a large amount of manually annotated training data.

\subsubsection{Part of Speech (PoS)}

A Part-of-Speech (PoS) is a category of words (or more generally, of lexical items) which have similar grammatical properties~\cite{martin2000speech}. Words that are assigned to the same part of speech generally display similar behavior in terms of syntax - they play similar roles within the grammatical structure of sentences - and sometimes in terms of morphology, in that they undergo inflection for similar properties. Commonly listed English parts of speech are noun, verb, adjective, adverb, pronoun, preposition, conjunction, interjection, and sometimes numeral, article or determiner.

\subsubsection{Keyword}

In corpus linguistics a keyword is a word which occurs in a text more often than we would expect to occur by chance alone~\cite{reitter2012system}. Keywords are calculated by carrying out a statistical test which compares the word frequencies in a text against their expected frequencies derived in a much larger corpus, which acts as a reference for general language use. To assist analysts filtering and indexing open data, it will be important to extract keywords from unstructured data such as Tweets text.

\subsubsection{Synonym}

A synonym is a word or phrase that means exactly or nearly the same as another word or phrase in the same language~\cite{henriksson2014synonym}. Words that are synonyms are said to be synonymous, and the state of being a synonym is called synonymy. An example of synonyms is the words begin, start, and commence. Words can be synonymous when meant in certain contexts, even if they are not synonymous within all contexts. For example, if we talk about a long time or an extended time, long and extended are synonymous within that context. Synonyms with exact interchangeability share a seme  or denotational sememe, whereas those with inexactly similar meanings share a broader denotational or connotational sememe and thus overlap within a semantic field~\cite{thomas2016towards}.

While analyzing the open data, it is important to extract the synonyms for the keywords and consider them in the analysis steps. For example, sometimes two Tweets can be related if we include the synonyms of the keywords in the analysis: instead of only focusing on the exact keyword match. It is important as the synonym can be a word or phrase that means exactly or nearly the same as another word or phrase in the Tweets.

\subsubsection{Stem}

A stem is a form to which affixes can be attached~\cite{sampson2005language}. For example, the word friendships contains the stem friend, to which the derivational suffix -ship is attached to form a new stem friendship, to which the inflectional suffix $-s$ is attached. To assist analysts understand and analyze the textual context, it will be important to extract derived form of the words in the text. For example, consider an analyst who is interested to identify the tweets that are related to health. Considering the keyword `health', using the Stem service, it is possible to identify derived forms such as healthy, healthier, healthiest, healthful, healthfully, healthfulness, etc; and more accurately identify the tweets that are related to health.

\subsubsection{Information Extraction from a URL}

A Uniform Resource Locator (URL), commonly informally termed a Web address is a reference to a Web resource that specifies its location on a computer network and a mechanism for retrieving it. Considering a Tweet that contains a URL link, it is possible to extract various types of information including: Web page title, paragraphs, sentences, keywords, phrases, and named entities. For example, consider a Tweet which contains URL links. It is possible to extract further information from the link content and use them to analyze the Tweets.

\subsection{Linking Services}

\subsubsection{Knowledge Bases}

While extracting various features (e.g. named entities, keywords, synonyms, and stems) from text, it is important to go one step further and link the extracted information items into the entities in the existing Knowledge Graphs (e.g. Google KG\footnote{https://developers.google.com/knowledge-graph/} and Wikidata\footnote{https://www.wikidata.org/}). For example, consider that we have extracted `M. Turnbull' from a tweet text. It is possible to identify a similar entity (e.g. `Malcolm Turnbull'\footnote{https://en.wikipedia.org/wiki/Malcolm\_Turnbull}) in the Wikidata knowledge base. As discussed earlier, the similarity API supports several function such as Jaro, Soundex, QGram, Jaccard and more. For this pair, the Jaro function returns 0.74 and the Soundex function returns 1.
To achieve this, we have leveraged the Google KG and Wikidata APIs to link the extracted entities from the text to the concepts and entities in these knowledge bases. For example, the Google API call will return a JSON file which may contain the url to wikipedia\footnote{https://en.wikipedia.org/}.

\subsubsection{Similarity}

\begin{figure} [t]
\centering
\includegraphics[width=0.8\textwidth]{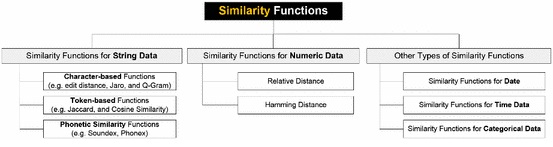}
\caption{Different categories of similarity functions~\cite{CDCR1}.}
\label{fig:similarity}
\end{figure}

Approximate data matching usually relies on the use of a similarity function, where a similarity function $f(v_1, v_2) \rightarrow s$ can be used to assign a score s to a pair of data values $v_1$ and $v_2$. These values are considered to be representing the same real world object if s is greater than a given threshold $t$.
In the last four decades, a large number of similarity functions have been proposed in different research communities, such as statistics, artificial intelligence, databases, and information retrieval. They have been developed for specific data types (e.g., string, numeric, or image) or usage purposes (e.g., typographical error checking or phonetic similarity detection). For example, they are used for comparing strings (e.g., edit distance and its variations, Jaccard similarity, and tf/idf based cosine functions), for numeric values (e.g., Hamming distance and relative distance), for phonetic encoding~\cite{CDCR1} (e.g., Soundex and NYSIIS), for images (e.g., Earth Mover Distance), and so on. The functions can be categorized as shown in Figure~\ref{fig:similarity}P, which illustrates different categories of similarity functions along with existing algorithms for each category~\cite{CDCR1,cheatham2013string,pinto2012soundex}.

\subsection{Classification Services}

Classification~\cite{jajuga2012classification,lin2014similarity,feldman2013techniques,bhardwaj2012data} is a data mining function that assigns items in a collection to target categories or classes. The goal of classification is to accurately predict the target class for each case in the data. For example, a classification model could be used to identify loan applicants as low, medium, or high credit risks. A classification task begins with a dataset in which the class assignments are known. For example, a classification model that predicts credit risk could be developed based on observed data for many loan applicants over a period of time. In addition to the historical credit rating, the data might track employment history, home ownership or rental, years of residence, number and type of investments, and so on. Credit ratings would be the target, the other attributes would be the predictors, and the data for each customer would constitute a case~\cite{hariharan2012discriminative}.

Classifications are discrete and do not imply order. Continuous, floating-point values would indicate a numerical, rather than a categorical target. A predictive model with a numerical target uses a regression algorithm, not a classification algorithm. The simplest type of classification problem is binary classification. In binary classification, the target attribute has only two possible values: for example, high credit-rating or low credit-rating. Multiclass targets have more than two values: for example, low, medium, high, or unknown credit rating. In the model build (training) process, a classification algorithm finds relationships between the values of the predictors and the values of the target. Different classification algorithms use different techniques for finding relationships. These relationships are summarized in a model, which can then be applied to a different data set in which the class assignments are unknown~\cite{andrews2012model}.

Classification models are tested by comparing the predicted values to known target values in a set of test data. The historical data for a classification project is typically divided into two data sets: one for building the model; the other for testing the model. Scoring a classification model, results in class assignments and probabilities for each case. For example, a model that classifies customers as low, medium, or high value would also predict the probability of each classification for each customer. Classification has many applications in customer segmentation, business modeling, marketing, credit analysis, and biomedical and drug response modeling~\cite{CDCR1,owen2012mahout}.

\subsection{Indexing Services}

For the developers, it is important to exposes the power of Elasticsearch~\cite{gormley2015elasticsearch} without the operational burden of managing it themselves. For example, it is important to automatically index entities and keywords for powerful, real-time Lucene\footnote{https://lucene.apache.org/} queries, e.g. while dealing with open data such as Twitter where every second, on average, around 6,000 tweets are tweeted on Twitter, which corresponds to over 350,000 tweets sent per minute, 500 million tweets per day and around 200 billion tweets per year.
In particular, an index is a data structure that improves the speed of data retrieval operations on data files at the cost of additional writes and storage space to maintain the index data structure. There are different indexing techniques to support structured (indexes in databases) and unstructured (keyword indexing for text files) data files.
For example, Apache Lucene  is a high-performance, full-featured text search engine. It is a technology suitable for nearly any application that requires full-text search, especially cross-platform. The API leverages Lucene and is able to get a set of keywords as an input and index the text files using these keywords. If no keywords provided, then the Lucene will automatically generate keywords by removing auxiliary verb, articles etc.

\subsection{Data and Meta-data Organization Services}

For the developers, it is important to easily interact with data layer: creating, reading, updating, deleting and querying data for Relational and NoSQL (e.g. Graph, Document and Key-Value stores) databases. This task is challenging and requires various skills and experiences: How to store key-value information, JSON documents, log data, and/or social graphs? How do we choose which DB to use? As on ongoing/future work, we are developing and API (i.e. CoreDB) to assist developers with the above mentioned challenges. We are focusing on manage multiple database technologies and weave them together at the application layer, and make this service accessible through a single
REST API.

\section{Motivating Scenario: Curating Twitter Data}
\label{MotivatingScenario}

\begin{figure} [t]
\centering
\includegraphics[width=0.3\textwidth]{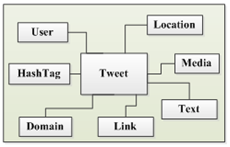}
\caption{Main artifatcs in Twitter.}
\label{fig:twitterschema}
\end{figure}

Large amount of (potentially related) data is being generated daily through many open and private data sources. For example, every second, on average, around 6,000 tweets are tweeted on Twitter , which corresponds to over 350,000 tweets sent per minute, 500 million tweets per day and around 200 billion tweets per year. As another example, we have more than 1.49 billion monthly active Facebook users who generate more than 4.5 billion likes daily . In particular, this data is large scale, never ending, and ever changing, arriving in batches at irregular time intervals. In this context, it will be vital to ingest this continuous-based data, persist it in database systems, extract features from it, summarize it (e.g. entity and keyword summaries), and make it easily available for the analysts to analyze the data and to gain value and insight from it. These tasks are challenging as Open Data sources generate complex, unstructured raw data at a high rate, resulting in many challenges to ingest, store, index, summarize and analyze such data efficiently. Consider Twitter as a motivating example.
%
%Machine learning techniques can be used to construct the schema for an open data source. Expert domain analysts can enrich this schema with more detailed features. Data Ingestion and Access services can be used as the bridge between a Data Source (e.g. Twitter Streaming Data) and the Data Lake (e.g. Twitter extracted folder).
Figure~\ref{fig:twitterschema} illustrates the main artifacts in Twitter. Figure~\ref{fig:twitterextract} illustrates the items that we can extract from Tweet Text. Figures~\ref{fig:twitterextract2} illustrate the items that we can extract from Twitter User artifact. For example, we extract named-entities, keywords, and synonyms from the Tweet text, from the content of the links mentioned in the Tweet text, and also from the user profile description of a person who twitted the Tweet.

\begin{figure}
\centering
\includegraphics[width=0.6\textwidth]{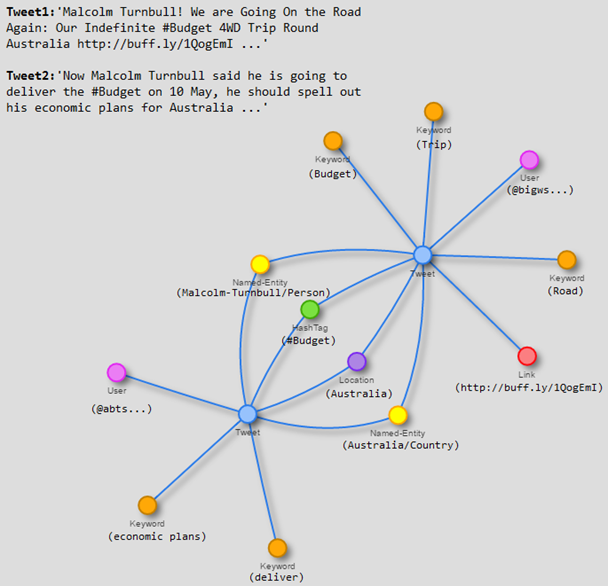}
\caption{Use extracted features from Twitter to link related Tweets.}
\label{fig:tweetGraph}
\end{figure}

\begin{figure}
\centering
\includegraphics[width=0.8\textwidth]{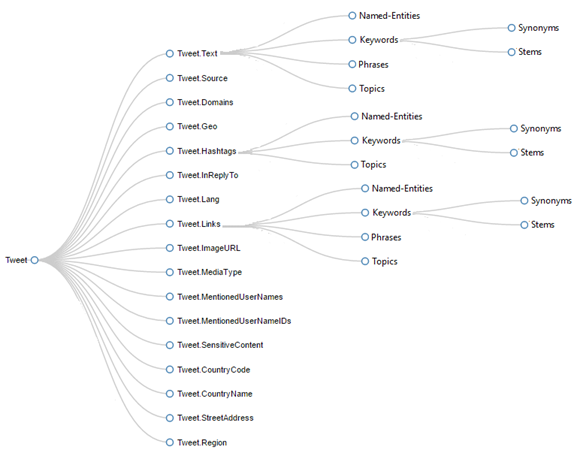}
\caption{Features that we can extract from Twitter Text.}
\label{fig:twitterextract}
\end{figure}

\begin{figure}
\centering
\includegraphics[width=0.8\textwidth]{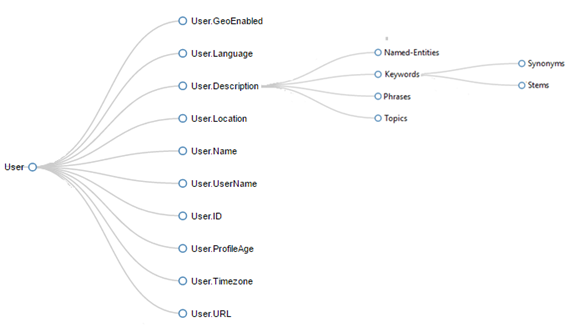}
\caption{Features that we can extract from Twitter Text.}
\label{fig:twitterextract2}
\end{figure}

\newpage

Extracting all these features will be a great asset to summarize the large number of Tweets. For example, `entity summaries' of Tweets containing the same named entity such as a person or organization; and "keyword summaries" of Tweets containing similar keywords. We may then analyze these related Tweets to get valuable insights from the Twitter open data. For example, consider Figure~\ref{fig:tweetGraph} where two real tweets have been illustrated. Using extracted folders, it is possible to extract information (e.g. named entities, keywords, and hashtags) from the Tweets and use them to generate a graph where nodes are the main artifacts and extracted information are the relationships among them. As illustrated in this figure, the Tweets are linked through named entities and hashtags and this will generate an interesting graph which reveals the hidden information among the nodes in the graph: for example it is possible to see the path (transitive relationships among the nodes and edges) between user1 and user2 which in turn reveals that these two users are interested in the same topics, and consequently may be part of some hidden micro-networks.

\newpage

\Huge
\begin{center}
Appendix
\end{center}
\LARGE

\begin{center}
Curation API's Documentation
\end{center}

\normalsize

\newpage

\section{Appendix: Named Entity Recognition API}

\subsection{Introduction}
Named Entity Recognition (NER) is the main component for every information extraction systems. NER system labels sequences of words in a text which are the names of things, such as person and company names, or gene and protein names. In particular, named entity extraction is the process of categorizing a text into different groups such as the name of persons, product, cities, countries, drugs, organizations, locations. NER systems receive a block of raw data, and return an annotated block of text, which contains a list of entities.
\subsection{API Specification}
\begin{figure}[h]
\centering
\includegraphics[width=0.75\textwidth]{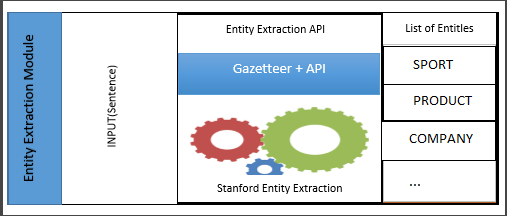}
\caption{Class diagram of Named Entity Extraction API}
\label{fig:classent}
\end{figure}

Named Entity Extraction API created on top of \emph{\textbf{S}tanford \textbf{N}amed \textbf{E}ntity \textbf{R}ecognizer} (SNER). SNER is a Java implementation of NER system and labels a block of text as the names of things, such as person, location, product, company and etc. The API not only extracts the entities provided as a part of NER system, but also contains a variety of methods for extracting other types of entities including, cities, products, job titles, medicines, sports.  It uses a dictionary (Gazetteer) with more than 30000 entities and mentions for improving the quality of named entity recognition. Also, the API, contains a set of methods for extracting the IP Address and Email Address using regular expressions. Figure ~\ref{fig:classent} shows the structure of API.

\subsection{API Implementation}
The API contains a set of libraries for extracting the entities of a file or a sentence namely: "ExtractEntiyFile", "ExtractEntitySentence" and "RegexClass". \emph{ExtractEntityFile} library extracts the entities of a file and \emph{ExtractEntitySentence} library extracts the entities from a sentence and \emph{RegexClass} library is a regular expression based library for extracting the email and IP-Addresses. Tables~\ref{tbl:regex} and~\ref{tbl:entity} show the methods provided as part of the API.

\begin{table}[h]
\centering

\begin{tabular}{||c c||}
\hline
 Method Name & Description\\
\hline\hline
ExtractEmailAddressFile & Returns Email addresses exist in file\\
\hline
ExtractEmailAddressSentence & Returns Email addresses exist in a sentence\\
\hline
ExtractIPAddressSentence & Returns IP addresses exist in a sentence\\
\hline
ExtractIPAddressFile & Returns IP addresses exist in file\\
\hline
\end{tabular}
\caption{Structure of "RegexClass" Library }
\label{tbl:regex}
\end{table}

\newpage
\begin{table}[h]
\centering

 \begin{tabular}{||c c||}
 \hline
 Method Name &  Description \\
 \hline\hline
ExtractFileNamedEntities & list of named entities in a file\\
\hline
ExtractSentenceNamedEntities & list of named entities in a sentence\\
\hline
ExtractContinent& Continent\\
\hline
ExtractOrganization & Organizations\\
\hline
ExtractCity & Name of Cities\\
\hline
ExtractCountry &Countries\\
\hline
ExtractCompany & Companies\\
\hline
ExtractPerson & Persons\\
\hline
ExtractLocation &	Locations\\
\hline
ExtractMoney &  Money\\
\hline
ExtractDate &Dates\\
\hline
ExtractDrug &Medicine\\
\hline
ExtractSport & Sport\\
\hline
ExtractHoliday &Holiday\\
\hline
ExtractNaturalDisaster & Natural Disaster\\
\hline
ExtractProduct & Product\\
\hline
ExtractOperatingSystem &  Operating System\\
\hline
ExtractSportEvent & sports events\\
\hline
ExtractState &list of States\\
\hline
ExtractDegree &  Degrees\\
\hline
ExtractMedia & Names of media\\
 \hline
 ExtractDisease & Name of Diseases\\
 \hline
\end{tabular}
\caption{Structure of ExtractEntityFile and ExtractEntitySentence libraries}
\label{tbl:entity}
\end{table}

\paragraph{Notice:}
"ExtractFileNamedEntites" method is available in the ExtractEntityFile library and "ExtractSentenceNamedEntity" is available in the ExtractEntitySentence library. Other methods are provided in both libraries. \\

\subsubsection{Gazetteer}
Gazetteer is responsible for categorizing the type and scope of the information presented. Gazetteer is a dictionary and it typically contains information concerning the geographical makeup of a country, region or continent as well as social statistics and physical features such as mountains, waterways or roads and etc.  The ExtractNamedEntity API uses a gazetteer for extracting the entities, which are not provided as a part of SNER system \cite{finkel2005incorporating}. Therefore, without the gazetteer, it is only able to recognize the location, organization, person, money, number and date entities. For finding other entities, the API uses a file named data.txt (located in the root of project).

\subsubsection{Creating Gazetteer}
\begin{figure}[h]
\centering
\includegraphics[width=1.0\textwidth]{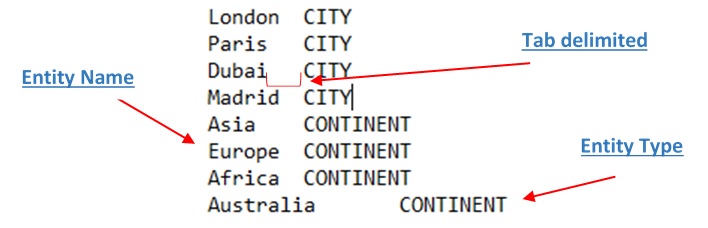}
\caption{Sample Gazetteer file}
\label{fig:gazetteer}
\end{figure}

Currently, The ExtractNamedEntity API uses a gazetteer, which contains more than 30000 entities. The gazetteer file is located at the root of project and named "data.txt". The file is tab delimited, and contains two columns. Entity-name and Entity-Type. In order to modify or create a new file follow the following steps. Figure ~\ref{fig:gazetteer} shows a sample gazetteer file.

\begin{enumerate}
 \item Open/Create a file, named data.txt in the root of your project.
 \item Add the entities and their types in a separate line.
 \item Save and close the file (the file name must be data.txt).
\end{enumerate}

\begin{figure}
\centering
\includegraphics[width=0.95\textwidth]{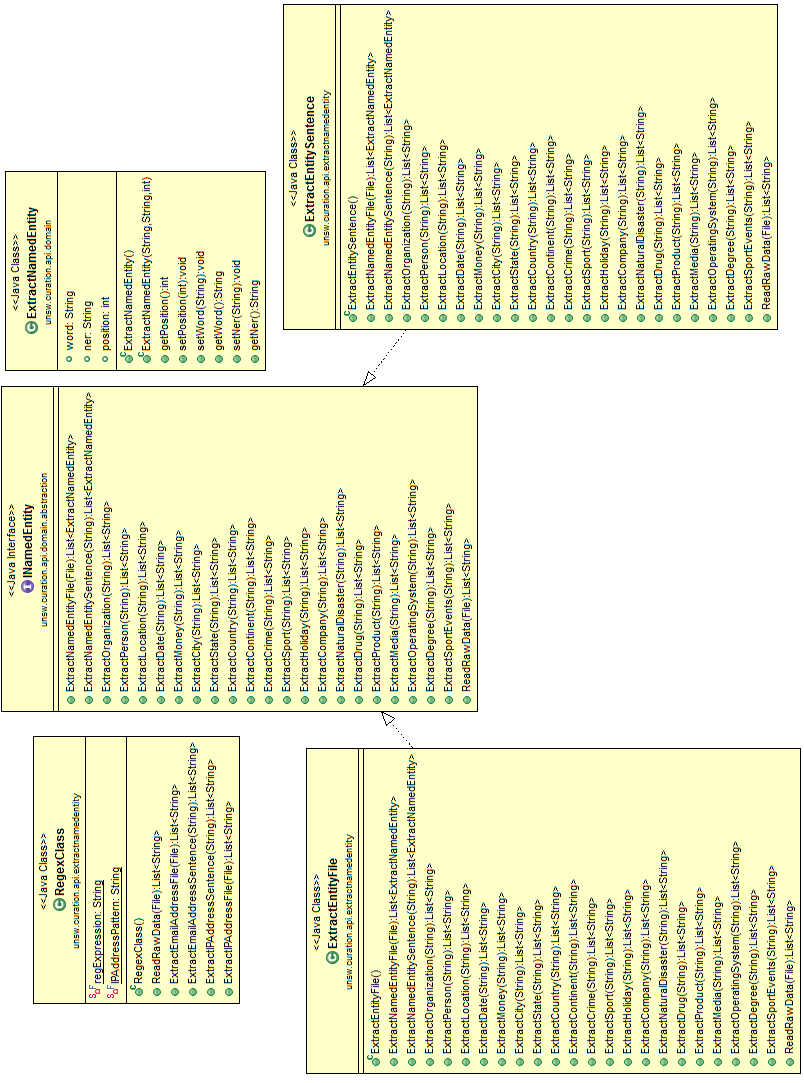}
\caption {Sample class diagram of ExtractNamedEntity API}
\label{fig:ExtractNamedEntity_ClassDiag}
\end{figure}

In the gazetteer file each entity must have a specific entity type. For example, if London is a city we ought to set the "London" as the entity and "CITY" as the entity type. Table~\ref{tbl:gaztteer_s} shows a guideline for setting the entity and entity types in the gazetteer file. Also figure~\ref{fig:ExtractNamedEntity_ClassDiag} shows the class diagram of the API.

\begin{table} [h]
\centering
 \begin{tabular}{||c c c||}
 \hline
 Method Name & Entity type & Sample\\
  \hline\hline
Name of City&	CITY&	Paris        CITY\\
  \hline
Name of Continent&	continent&	Asia        continent\\
  \hline
Name of Country&	COUNTRY&	USA        COUNTRY\\
  \hline
Name of Drug&	DRUG&	Acarbose        DRUG\\
  \hline
Name of Company&	COMPANY&	Asus        COMPANY\\
  \hline
Name of Crime&	Crime&	Larceny	Crime\\
  \hline
Name of Sport&	Sport&	Archery 	Sport\\
  \hline
Name of Holiday&	holiday&	Christmas	holiday\\
  \hline
Name of Product&	product&	Bag	Product\\
  \hline
Name of Natural disaster&	disaster&	Flood          disaster\\
  \hline
Name of Operating System&	os&	Ubuntu            os\\
  \hline
Name of Sport Event&	sportev&	Asian Cup	sportev\\
  \hline
Name of Geographic Feature&	geo&	Cliff       	geo\\
  \hline
Region	Region&	Region&	Suncoast region\\
  \hline
Name of State&	state&	New South Wales   State\\
  \hline
Name of a degree&	degree&	Senior Lecturer   degree\\
  \hline
\end{tabular}
\label{tbl:gaztteer_s}
\caption{Structure of Gazetteer file}
\end{table}

\paragraph{Notice:}
The methods provided in table 4.2 receive a sentence or a file path (String) and return a string of entities. However, "ExtractNamedEntityFile" and "ExtractNamedEntitySentence" methods return a list, type of "ExtractNamedEntity" object, which contains two properties; \emph{Word} and \emph{Ner}. \emph{Word} refers to the entity and \emph{Ner} refers to the entity type. For processing the output of \emph{ExtractNamedEntitySentence} and \emph{ExtractNamedEntityFile} methods, create a list, type of "ExtractNamedEntity" and iterate the list using the \emph{"for"} loop statement.
\newpage
\subsection{Evaluation}
\begin{figure}
\centering
\includegraphics[width=0.8\textwidth]{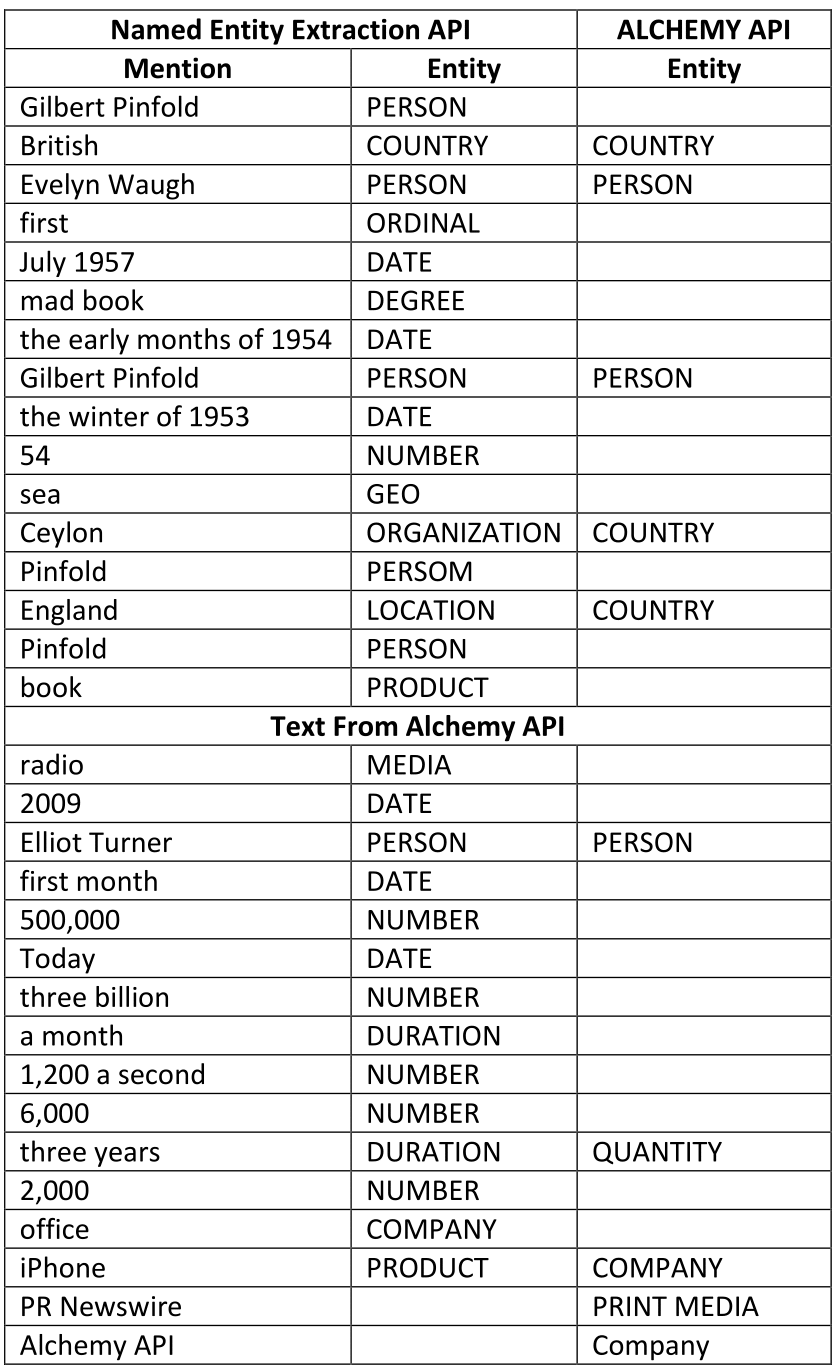}
\captionof{table}{Results obtained from Alchemy API and Named Entity Extraction API}
\label{fig:result_ent}
\end{figure}

We have evaluated the performance of Named Entity Extraction API and Alchemy API. For evaluation we used two source of data, one from Wikipedia web site and the other one was the text available as test data in the Alchemy API web site. the extracted result proved that our API has acceptable result and could improve the performance of SNER system. Table ~\ref{fig:result_ent} shows the performance of Named Entity Extraction API and the Alchemy API.

\subsection{User Guide}
In this section we explain the steps need to extract named entities using the API. Follow the following steps for extracting the named entities.
\begin{enumerate}
\item Open the project in Eclipse IDE (File -\textgreater Import -\textgreater Maven -\textgreater Existing Maven Projects -\textgreater Next -\textgreater click browse and select the folder that is the root of the maven project -\textgreater click ok).
\item Create a new class and add the following method to your class -\textgreater public static void main(String [] args).
\item Import unsw.curation.api.domain.ExtractNamedEntity and \\ unsw.curation.api.extr actnamedentity.ExtractEntityFile (for extracting entities from file) and unsw.curation.api.extractnamedentity.ExtractEntitySentence (for extracting entities of a sentence).
\item Create an instance of "ExtractEntityFile" or "ExtractEntitySentence".
\item Call one of the provided methods.
\item Iterate the result using "for" loop statement.
\end{enumerate}
\subsubsection{Extracting Entities of a File}
In order to extract the entities of a file, create an instance of "ExtractEntityFile" library and call the "ExtractNamedEntityFile" method. Following shows the sample code.
\newline \textbf{Example:}\newline

\begin{center}
 \begin{tabular}{c}
  \setlength\fboxsep{0pt}
  \setlength\fboxrule{0.0pt}
  \fbox{\includegraphics[scale=0.45]{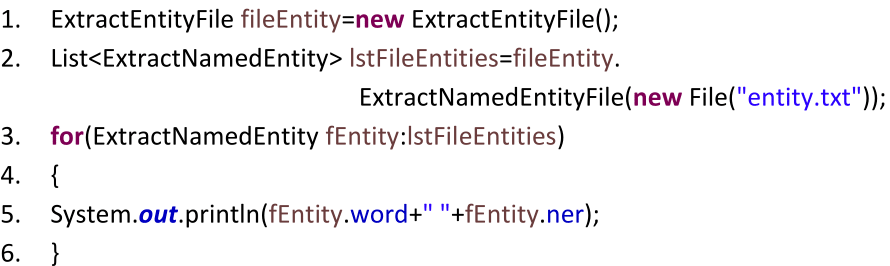}}
\end{tabular}
\end{center}

\subsubsection{Extracting Entities of a Sentence}
In order to extract the entities of a sentence, create an instance of "ExtractEntitySentence" library and call the "ExtractNamedEntitySentence" method. Following shows the sample code.\newline \textbf{Example:}\newline
\begin{center}
 \begin{tabular}{c}
  \setlength\fboxsep{0pt}
  \setlength\fboxrule{0.0pt}
  \fbox{\includegraphics[scale=0.45]{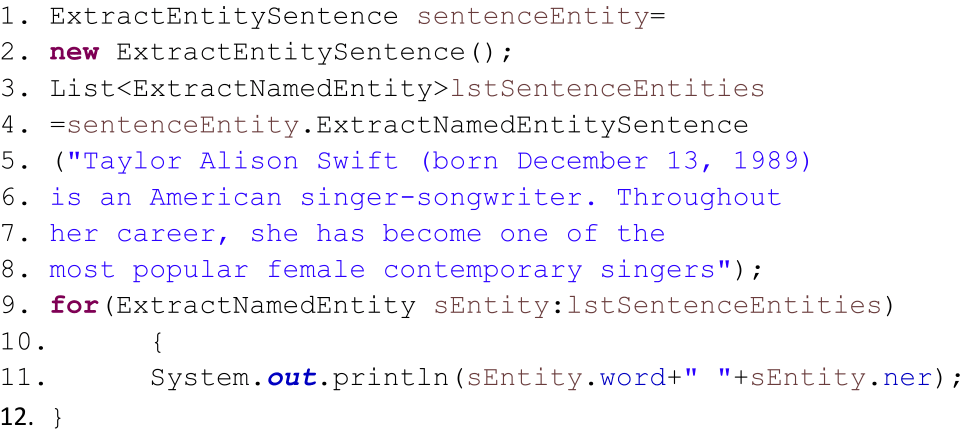}}
\end{tabular}
\end{center}

\subsubsection{Extracting a specific entity type}
"ExtractNamedEntity" API contains a set of methods for extracting a specific entity type (city, person, product, company). In order to extract these entity types create an instance of "ExtractEntitySentence" or "ExtractEntityFile" class libraries and call the appropriate methods. Following shows a sample code for extracting a specific entity from a file or a sentence.\\ \textbf{Example:}\newline

\begin{center}
 \begin{tabular}{c}
  \setlength\fboxsep{5pt}
  \setlength\fboxrule{0pt}
  \fbox{\includegraphics[scale=0.40]{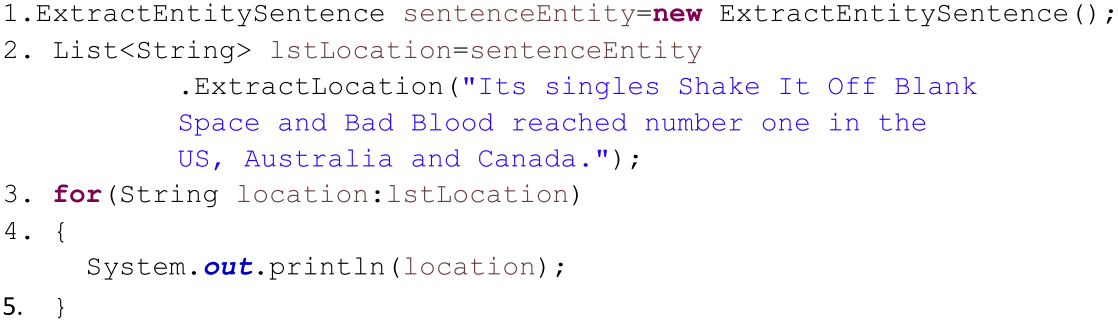}}
\end{tabular}
\end{center}

\begin{center}
 \begin{tabular}{c}
  \setlength\fboxsep{5pt}
  \setlength\fboxrule{0pt}
  \fbox{\includegraphics[scale=0.40]{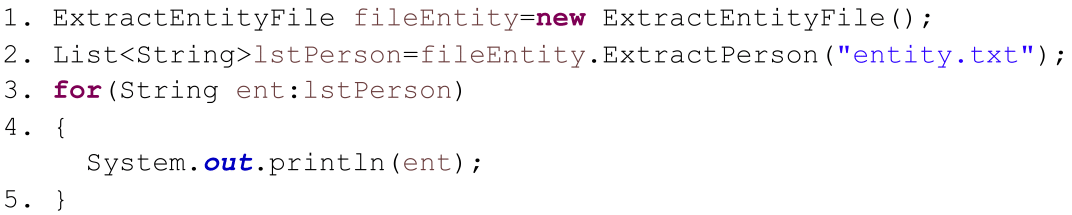}}
\end{tabular}
\end{center}

\newpage

\section{Appendix: Part Of Speech Extraction API}

\subsection{Introduction}
\textbf{P}art \textbf{O}f \textbf{S}peech (POS) Tagging is the task of labeling a sequence of tokens as matching to a specific part of speech \footnote{\path{https://en.wikipedia.org/wiki/Part-of-speech_tagging}}. In particular, part of speech tagger receives a text and assigns parts of speech to each word (and other token), such as noun, verb, adjective. POS tagging is not possible by just having a list of words and their POS, because some words have different POS at different times.
\subsection{API Specification}
\begin{figure} [h]
\centering
\includegraphics[width=0.75\textwidth]{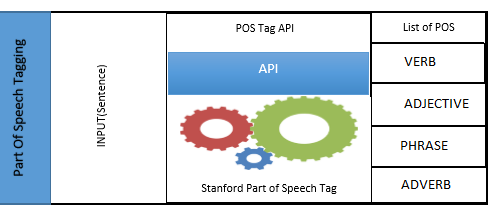}
\caption{Structure of POS Tag Extraction API.}
\label{fig:pos_structure}
\end{figure}

We have implemented an API for POS tagging based on \textbf{S}tanford \textbf{P}OS(SPOS) Tagger. The API contains a variety of methods for extracting the verbs, adjective, adverb, parse tree, quotation. Figure ~\ref{fig:pos_structure} shows the structure of the API.

\subsection{API Implementation}
\begin{figure} [h]
\centering
\includegraphics[width=0.6\textwidth]{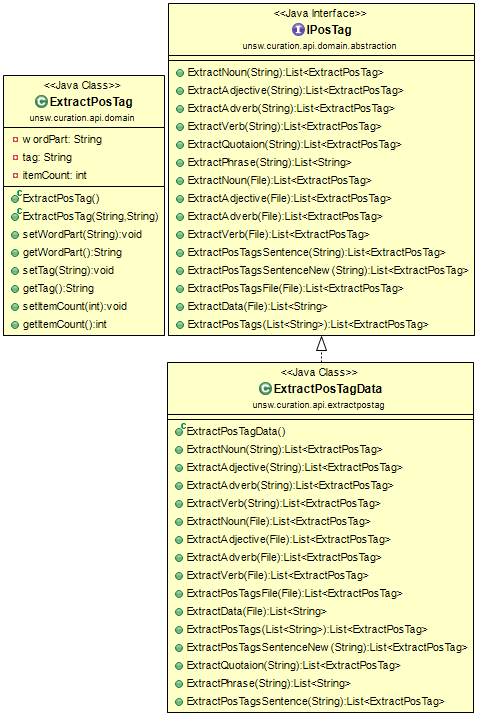}
\caption{Sample class diagram of ExtractPOSTag API}
\label{fig:pos_class}
\end{figure}

For extracting the POS of a given data the API contains a library named "ExtractPOSTagData", which extracts the POS tags of an input file or a sentence. Table ~\ref{tbl:pos_structure_library} shows the list of methods and their descriptions and figure ~\ref{fig:pos_class} shows the class diagram of the API.

\begin{center}

 \begin{tabular}{||c c||}

 \hline
 Method Name &  Description\\
  \hline\hline
ExtractVerb&	Extract Verbs\\
  \hline
ExtractNoun&	Extract Nouns\\
  \hline
ExtractAdjective& Extract Adjectives\\
  \hline
ExtractAdverb&	Extract Adverb\\
  \hline
ExtractPhrase&	Extract Noun and Verb Phrases\\
  \hline
ExtractQuotation&Extract POS Tag of texts between two Quotation\\
  \hline
ExtractPOSTags &  Extract all POS tags\\
\hline
\end{tabular}
\captionof{table}{Structure of Part Of Speech library}
\label{tbl:pos_structure_library}
\end{center}
The methods presented in table ~\ref{tbl:pos_structure_library},  receives an unannotated block of text and returns a collection of labeled tokens. ExtractPhrase method, contrary to the rest of methods returns a tree from noun phrases and verb phrases exists in the text.
\paragraph{Notice}
The methods provided in the API (Part of Speech Extraction) return a list of string. However, "ExtractPOSTags" method returns a list, type of \emph{ExtractPosTag} object, which contains two properties; \emph{WordPart} and \emph{Tag}. \emph{WordPart} is the token and \emph{Tag} is the POS of token. For processing the result, create a list type of \emph{ExtractPosTag} and iterate the list using the \emph{"for"} loop statment (for further detail refer to the user guide section).

\subsection{Evaluation}
\begin{figure} [h]
\centering
\includegraphics[width=0.45\textwidth]{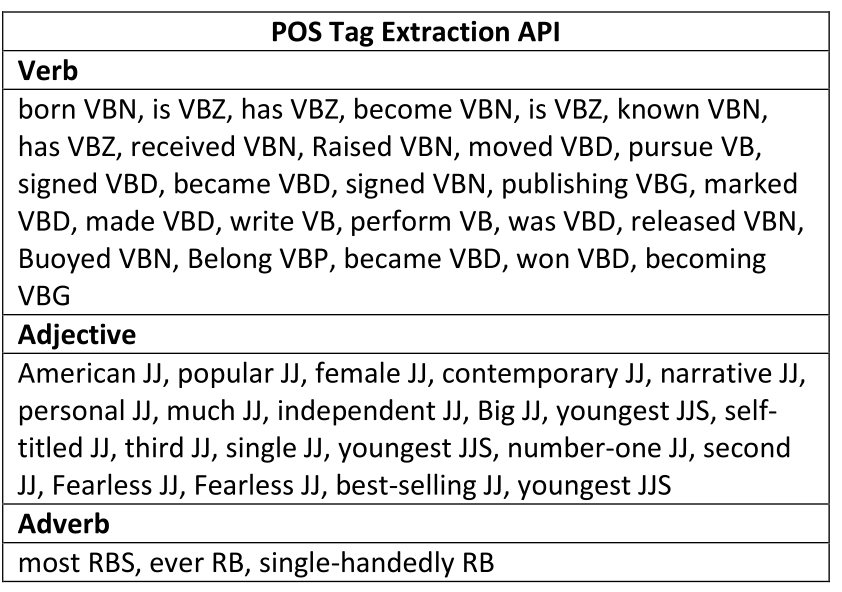}
\captionof{table}{The output result obtained from POS Tag Extraction API}
\label{tbl:pos_output}
\end{figure}

In this section we present the output that we have obtained by using the Part of Speech Extraction API. We extracte some text from Wikipedia and fed them as the input to the API. In order to use the API refer to Section 5.5 (User Guide). Also the sample codes, are available in the source code of project (unsw.curation.api.run package).  Table ~\ref{tbl:pos_output} shows a part of knowledge that can be extracted from the POS Tag Extraction API.

\subsection{User Guide}
In this section we explain the process of extracting the POS tags of a sentence or a file using the "ExtractPOStag" API. "ExtractPOSTagData" is the responsible library, for extracting the POS tags and contains all required methods for POS tagging. "ExtractPOSTagFile" and "ExtractPOSTagSentence" are the two methods, which return the POS Tags of a given sentence or a file; other methods, such as ExtractVerb, ExtractNoun (the complete list of methods are available in table ~\ref{tbl:pos_structure_library}) return a specific POS Tag from a text. In order to use the API, follow the following steps;
\begin{enumerate}
\item Import unsw.curation.api.domain.ExtractPosTag and unsw.curation.api.extr\\actPOSTagData packages.
\item Create an instance of ExtractPOSTagData.
\item For extracting a specific POS Tags call ExtractNoun, ExtractVerb,\\ ExtractAdverb, ExtractAdjective methods.
\item For extracting all possible POS tags in a text call ExtractPOSTagS\\entence (for Extracting the POS tags in sentence) and ExtractPOSTagFile (for extracting the POS tags in within a file).
\item Iterate the result using \emph{"for"} loop statement.
\end{enumerate}

\textbf{Example:} \newline
\begin{center}
 \begin{tabular}{c}
  \setlength\fboxsep{5pt}
  \setlength\fboxrule{0pt}
  \fbox{\includegraphics[scale=0.45]{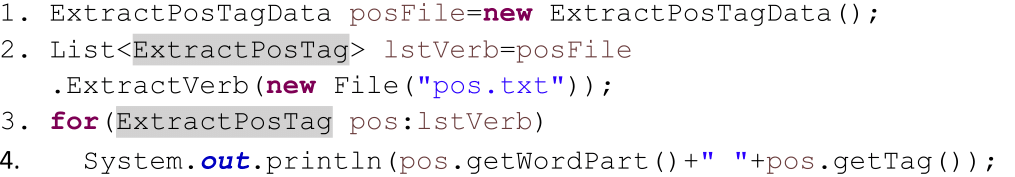}}
\end{tabular}
\end{center}
\textbf{Example:} \newline
\begin{center}
 \begin{tabular}{c}
  \setlength\fboxsep{5pt}
  \setlength\fboxrule{0pt}
  \fbox{\includegraphics[scale=0.45]{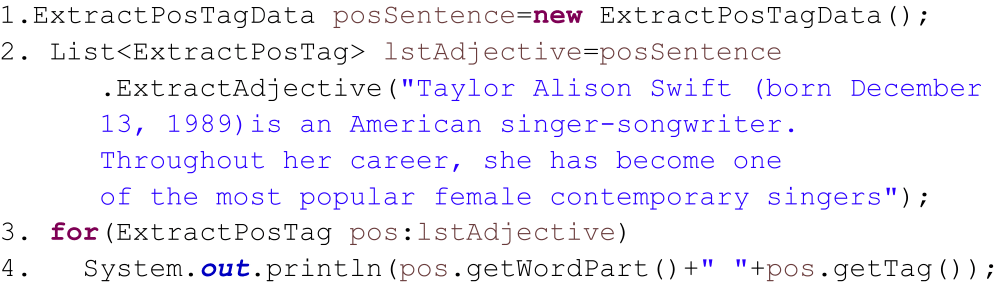}}
\end{tabular}
\end{center}

\newpage

\section{Appendix: Synonym API}

\subsection{Introduction}
Synonym API extracts the synsets of a token based on the WordNet dictionary. WordNet is a lexical database for the English language [1]. It groups English words into sets of synonyms called synsets and records a number of relations among these synonym sets or their members \footnote{\path{www.wikipedia.org}}.
WordNet is a dictionary\footnote{wordnet.princton.edu}, which collects similar words based on their synsets and extracts the relationship exist among the words. WordNet is used in a variety of applications, including word sense disambiguation, text classification, machine translation and information retrieval.
\subsection{API Specification}
\begin{figure} [h]
\centering
\includegraphics[width=0.70\textwidth]{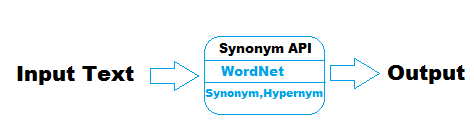}
\captionof{figure}{Structure of Synonym API}
\label{fig:syn_ext}
\end{figure}

Synonym API extracts synonym and hypernym of a token using WordNet \cite{miller1995wordnet} dictionary. The API is implemented on top of wordnet dictionary and contains a set of methods for improving the process of knowledge extraction from Wordnet. Figure ~\ref{fig:syn_ext} shows the structure of the API.

\subsection{API Implementation}
\begin{figure} [h]
\centering
\includegraphics[width=0.70\textwidth]{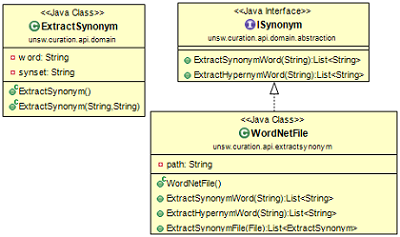}
\captionof{figure}{class diagram of synonym API}
\label{fig:syn}
\end{figure}

At the core of API is a library named "WordNetFile", which has the role of extracting synsets. Table ~\ref{tbl:synAPI} shows the methods and their descriptions. Also, figure ~\ref{fig:syn} shows the class diagram of the API.

\begin{table} [h]
\centering

 \begin{tabular}{||c c||}
 \hline
 Method Name &  Description\\
 \hline\hline
ExtractSynonymWord & returns synonyms of a word\\
 \hline
ExtractHypernymWord & returns hyponyms of a word\\
 \hline
ExtractSynonymSentence & returns synonyms of words in a sentence\\
 \hline
ExtractSynonymFile & returns synonyms of words in a file\\
 \hline
\end{tabular}
\caption{Structure of ExtractSynonymAPI}
\label{tbl:synAPI}
\end{table}

\paragraph{Notice:}
"ExtractSynonymSentence" and "ExtractSynonymFile" methods return a list, type of "ExtractSynonym",  which contains two properties; \emph{Word} and \emph{Synset}. \emph{word} contains the token and the \emph{Synset} contains the synonym or the hypernym of token. In order to process the output of the API, create a list type of "ExtractSynset" and iterate the list using the \emph{"for"} statement loop. (For further detail refer to the user guide section).
\subsection{Evaluation}
\begin{figure} [h]
\centering
\includegraphics[width=0.69\textwidth]{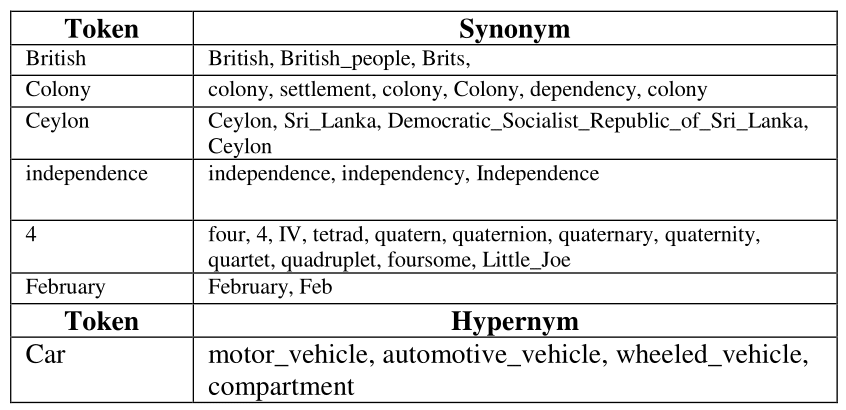}
\captionof {table}{The output obtained from Synonym Extraction API}
\label{fig:synoutput}
\end{figure}

In this section we present a part of output that obtained by extracting the synonym and hypernym of a Wikipedia text using the API. For demonstrating the performance of the API we have used the following sentence. Also for demonstrating the output of hypernym method, we simply used a token (car). The output of both methods are presented in table ~\ref{fig:synoutput}.
\begin{center}
The British Colony of Ceylon achieved independence on 4 February 1948, with an amended constitution taking effect on the same date. Independence was granted under the Ceylon Independence Act 1947.
\end{center}

\subsection{User Guide}
\emph{"WordNetFile"} is the library, placed at the core of API and contains a set of methods, including "ExtractSynonymSentence", "ExtractSynonymFile", "ExtractHypernymWord" and "ExtractSynonymWord" for extracting synonyms and hypernyms of an input text. Following shows the steps needs for using the API.

\begin{enumerate}
\item Open the project in Eclipse IDE (File -\textgreater Import -\textgreater Maven -\textgreater Existing Maven Projects -\textgreater Next -\textgreater click Browse and select the folder that is the root of the maven project -\textgreater click ok).
\item Create a new class and add the following method to your class -\textgreater public static void main(String [] args).
\item  Import unsw.curation.api.extractsynonym.WordNetFile to your project.
\item  Create an instance of WordNetFile (If you already haven't installed the wordnet dictionary, you need to set the dictionary path in the constructor of the library).
\item  Call methods provided in the library.

\end{enumerate}
\textbf{Example:}\newline
\begin{center}
 \begin{tabular}{c}
  \setlength\fboxsep{5pt}
  \setlength\fboxrule{0pt}
  \fbox{\includegraphics[scale=0.40]{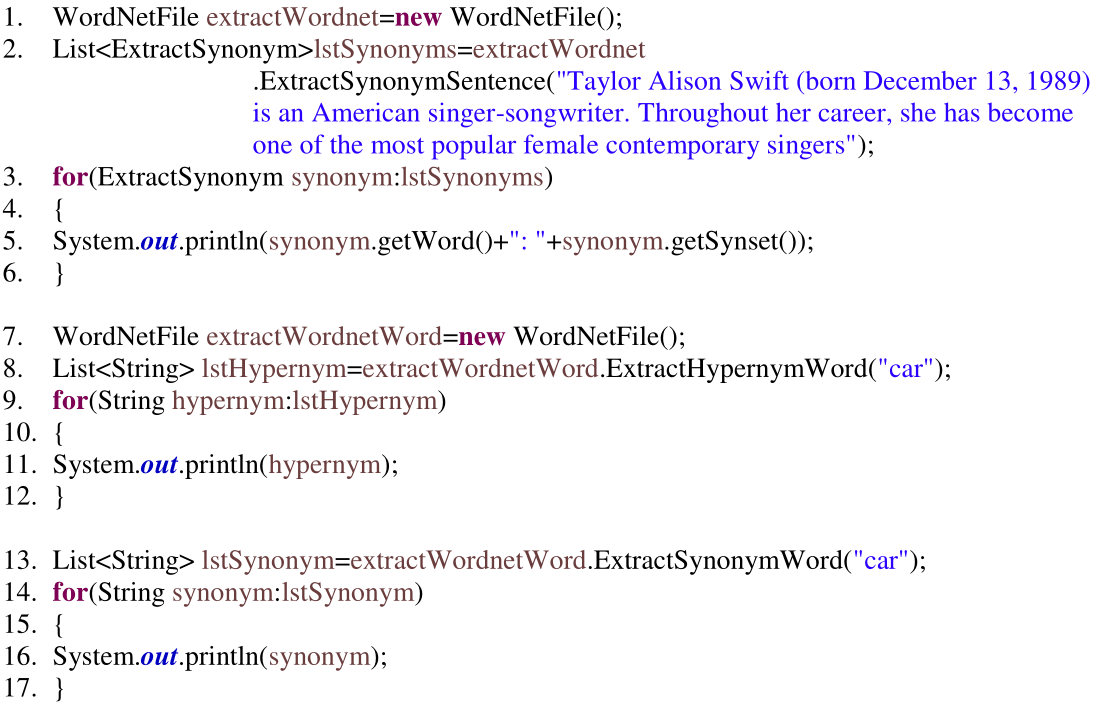}}
\end{tabular}
\end{center}
\paragraph{Notice:}
The API needs "WordNet" dictionary for extracting the synonyms and hyponyms. If \emph{WordNet} 2.1 already installed with the default settings \path{"C:\Program Files (x86)\WordNet\2.1\dict\"}, the API detects the existence of the dictionary, otherwise provide the \emph{WordNet} dictionary path while using the API.
\newpage
\section{Appendix: URL Extraction API}
\subsection{Introduction}
URL Extraction API is a library, for fetching and parsing html tags using JSOUP library. It provides a convenient API for extracting and manipulating data, using the best of DOM, CSS, and jquery-like methods. Jsoup implements the WHATWG HTML5 specification, and parses HTML to the same DOM as modern browsers do.
\subsection{API Specification}
\begin{figure}[h]
\centering
\includegraphics[width=0.70\textwidth]{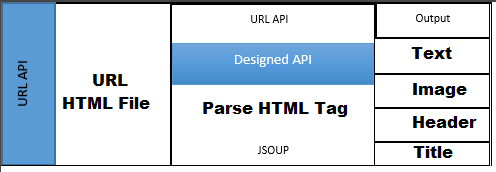}
\captionof{figure}{Structure of URL API}
\label{fig:url_struct}
\end{figure}

URL API implemented on top of JSOUP library and contains a set of methods to facilitate the process of parsing and fetching html tags. The API can parse both url's and html files. The Structure of API is presented in figure ~\ref{fig:url_struct}

\subsection {API Implementation}
\begin{figure}[h]
\centering
\includegraphics[width=0.65\textwidth]{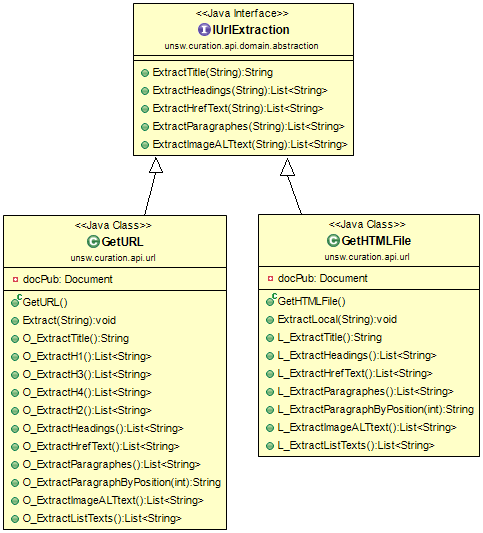}
\captionof{figure}{class diagram of URL API}
\label{fig:class_url}
\end{figure}

Processing input data is the first step in accessing the content of a html page.  URL API contains two libraries for parsing the html tags. The first one extracts the html tags from a url and the second one is suitable for parsing a local html documents in a file system. both libraries contain similar methods."ExtractTitle" method extracts the title of a html page. "ExtractHeading" and "ExtractHrefText" methods extract the text of heading (H tags) and the caption of images respectively. Finally "ExtractParagraph" method, extracts the content of \emph{P} tags. Table ~\ref{tbl:url_API} shows the methods and their descriptions and figure ~\ref{fig:class_url} show the class diagram of API.

\begin{center}

 \begin{tabular}{||c c||}
 \hline
 Method Name &  Description\\
  \hline\hline
  Extract(String URL)& parse a HTML document\\
  \hline
  ExtractTitle&Extract the title \\
  \hline
  ExtractHeading&Extract H1,H2,H3,H4 tags\\
  \hline
  ExtractHrefText&ExtractHref Tags\\
  \hline
  ExtractParagraphes&Extract P Tags\\
  \hline
  ExtractParagraphesByPosition(int Position)&Extract a specified P Tag\\
  \hline
  ExtractImageAltText&Extract Img Tags\\
  \hline
  ExtractListText &Extract UL tags\\
\hline
\end{tabular}
\captionof{table}{List of methods available in the ExtractURL API}
\label{tbl:url_API}
\end{center}

\paragraph{Notice:}
if an error occurs while fetching the data the API will throw an IO Exception.

\subsection{Evaluation}

\begin{center}
 \begin{tabular}{c}
  \setlength\fboxsep{0pt}
  \setlength\fboxrule{0.0pt}
  \fbox{\includegraphics[scale=0.45]{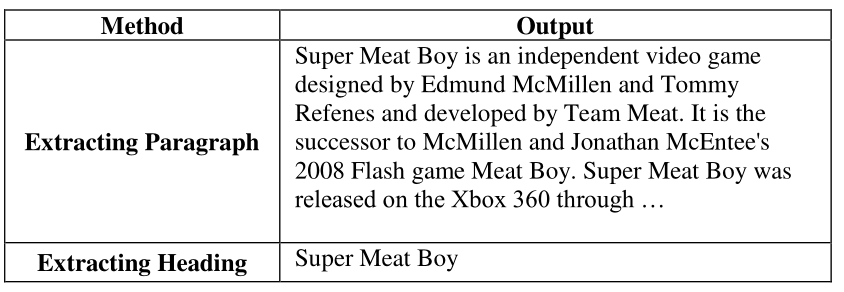}}
\end{tabular}
\captionof{table}{The output obtained from URL Extraction API}
\label{tbl:ext_API}
\end{center}

In this section we demonstrate a part of result that can be obtain from the API. We have extracted the content and Headings (content of H tags) of a url (a Wikipedia web page) using the "ExtractParagraph" and "ExtractHeading" methods. Also the sample code for using the API is available in the source code of project. Table~\ref{tbl:ext_API} shows the outputs obtained from URL Extraction API.

\subsection {User Guide:}
 The API contains several methods for parsing the html tags. The API receive a URL or a html file path and return the content of a given url or a file. In the first step, create an instance of GetURL or GetHTMLFile libraries (the first one is designed for fetching the html content of an url and the second one parse the content of a local html document). Then call one of the available methods. following shows the steps need for parsing the content of an HTML file or URL.
\begin{enumerate}
\item Open the project in Eclipse IDE (File -\textgreater Import -\textgreater Maven -\textgreater existing Maven Projects -\textgreater Next -\textgreater click browse and select the folder that is the root of the maven project -\textgreater click Ok).
\item Create a new class and add the following method to your class -\textgreater public static void main(String [] args).
\item import unsw.curation.api.url.GetUrl (fetching html tags of a url) or unsw.c\\uration.api.GetHtmlFile(parsing the html tags of a html document).
\item create an instance of GetUrl or GetHtmlFile libraries.
\item call Extract method (fetching the html content of a url) or ExtractLocal (setting the html file path).
\item call one of the methods provided in the libraries for extracting the content of a specific tags.
\item print or iterate the results.
\end{enumerate}
Following shows a sample code, which extracts the text of a Wikipedia page.\\

\textbf{Example:}\newline
\begin{center}
 \begin{tabular}{c}
  \setlength\fboxsep{5pt}
  \setlength\fboxrule{0pt}
  \fbox{\includegraphics[scale=0.50]{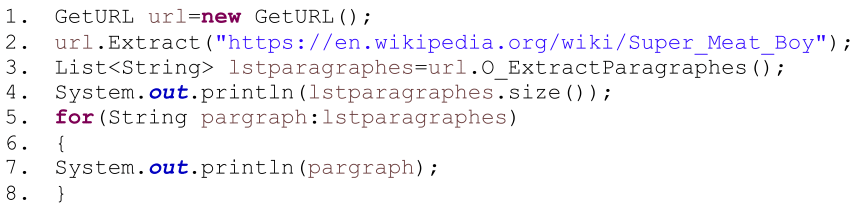}}
\end{tabular}
\end{center}

\newpage

\section{Appendix: Tokenization API}

\subsection{Introduction}

Tokenization is the task of chunking a stream of words into tokens and other meaningful symbols. Tokenization returns the keywords of a stream of words by removing the punctuations and stopwords.

\subsection{API specification}
The API uses LUCENE (Information Retrieval API) \footnote{\path{https://lucene.apache.org/core/}} for conducting the fast tokenization. Apache Lucene is a free and open-source information retrieval software library. It is supported by the Apache Software Foundation and is suitable for full text indexing and searching capability.
Lucene API can index different types of files, including PDFs, html, microsoft word, mind maps, and open document documents, as well as many others (except images), as long as their textual information can be extracted.

\subsection {API Implementation}
The API is designed to tokenize the content of a text or a file. At the core of the API is the "ExtractTokenclass" library, which is responsible for tokenizing a text. "ExtractSentenceToken" and "ExtractFileToken" are the methods, which receive a sentence or a file and return a string of comma delimited tokens. Also, "ExtractListToken" method, tokenize a collection of text placed in a list. Table ~\ref{tbl:tok_lib} shows the method available in the API and figure ~\ref{fig:tok_class} shows the class diagram.

\begin{center}
 \begin{tabular}{c}
  \setlength\fboxsep{5pt}
  \setlength\fboxrule{0pt}
  \fbox{\includegraphics[scale=0.55]{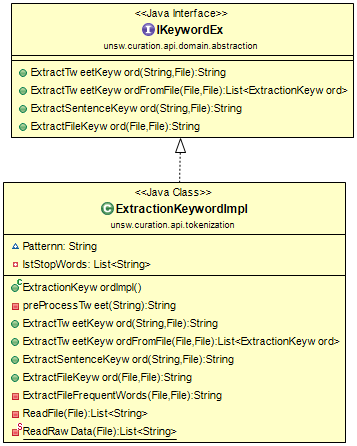}}

\end{tabular}
 \captionof{figure}{Class diagram of Tokenization API}
 \label{fig:tok_class}

\end{center}
\begin{center}

 \begin{tabular}{||c c||}
 \hline
 Method Name &  Description\\
  \hline\hline
  ExtractSentenceToken& Extract Tokens of a sentence\\
  \hline
  ExtractFileToken&Extract Tokens of a file\\
  \hline
  ExtractListToken&Extract tokens of a list\\
  \hline
\end{tabular}
\captionof{table}{Structure of ExtractTokenClass library}
\label{tbl:tok_lib}
\end{center}
\subsection{Evaluation}
We have omitted the output of tokenization API for brevity. However the sample code is provided in the source code of project (unsw.curation.api.run package). Also section 8.5 shows the steps needs to use the API.
\subsection{User Guide}
The API uses a list of stopwords and returns the keywords of a given text. Tokenization API uses LUCENE (Information Retrieval Library) for tokenizing data. Following code shows the process of extracting the keywords of an input sentence or a file.
\begin{enumerate}
\item Open the project in Eclipse IDE (File -\textgreater Import -\textgreater Maven -\textgreater Existing Maven Projects -\textgreater Next -\textgreater click Browse and select the folder that is the root of the maven project -\textgreater Click Ok)
\item Create a new class and add the following method to your class -\textgreater public static void main(String [] args)
\item  Import unsw.curation.api.tokenization package.
\item Create an instance of ExtractionKeywordImpl library.
\item use one of the methods provide in the library.
\end{enumerate}
\textbf{Example:}\newline
\begin{center}
 \begin{tabular}{c}
  \setlength\fboxsep{5pt}
  \setlength\fboxrule{0pt}
  \fbox{\includegraphics[scale=0.50]{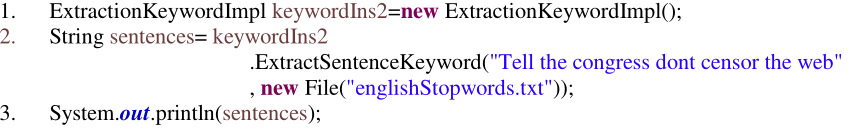}}
\end{tabular}
\end{center}
\textbf{Example:}\newline
\begin{center}
 \begin{tabular}{c}
  \setlength\fboxsep{5pt}
  \setlength\fboxrule{0pt}
  \fbox{\includegraphics[scale=0.50]{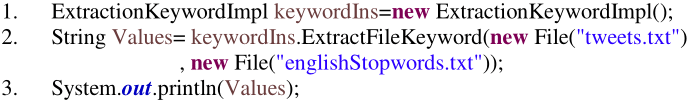}}
\end{tabular}
\end{center}

\newpage

\section{Appendix: Similarity API}

\subsection{Introduction}
Similarity API can be used for calculating the similarity of different values. The API contains several libraries, including cosine, tf-idf, jaccard, euclidean, leveneshtein, soundex and etc. Euclidean library calculates the similarity of numerical vectors. TF-IDF, Leveneshtein, soundex and qgram libraries compute the similarity of textual values. Cosine and Jaccard libraries compute the similarity of numerical vectors and textual values. Figure ~\ref{fig:sim_struct} shows the structure of Similarity API.
\begin{center}

 \begin{tabular}{c}
  \setlength\fboxsep{5pt}
  \setlength\fboxrule{0pt}
  \fbox{\includegraphics[scale=0.8]{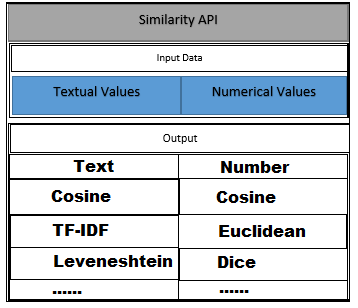}}
\end{tabular}
\captionof{figure}{Structure of Similarity API}
\label{fig:sim_struct}
\end{center}
\subsection {API Specification}
Similarity API is a rich API for computing the similarity of texts and numerical values. In this section we briefly introduce the similarity metrics exist in the API.
\textbf{Cosine} similarity is a famous similarity metric and calculates the cosine of the angle between two vectors. Cosine similarity is widely used metric in field of information retrieval and indicate the measure of similarity between two documents.
\textbf{Jaccard} coefficient is a similarity metric for calculating the similarity and diversity of sample sets.
TF-IDF, is a numerical statistic and indicate how important a word is to a document in a collection. \textbf{Tf-idf} metric mostly used as a weighting factor in information retrieval and textmining. In particular, TF-IDF weight consists of two parts. the first part computes the normalized Term Frequency (TF), the number of times a word appears in a document, divided by the total number of words in that document; the second part is the Inverse Document Frequency (IDF), computed as the logarithm of the number of the documents in the corpus divided by the number of documents where the specific term appears.
The Euclidean distance is the distance between two points. \textbf{Euclidean} distance examines the root of square difference between coordinates of a pair of objects.
\textbf{Levenshtein} distance is a string metric for measuring the difference between two token. The Levenshtein distance between two words is the number of edits (insertions, deletions, substitutions, transposition) require to change one word into an other. Figure ~\ref{fig:sim_class} shows a part of Similarity API class diagram.

\begin{center}
  \setlength\fboxsep{5pt}
  \setlength\fboxrule{0pt}
  \fbox{\includegraphics[scale=0.57]{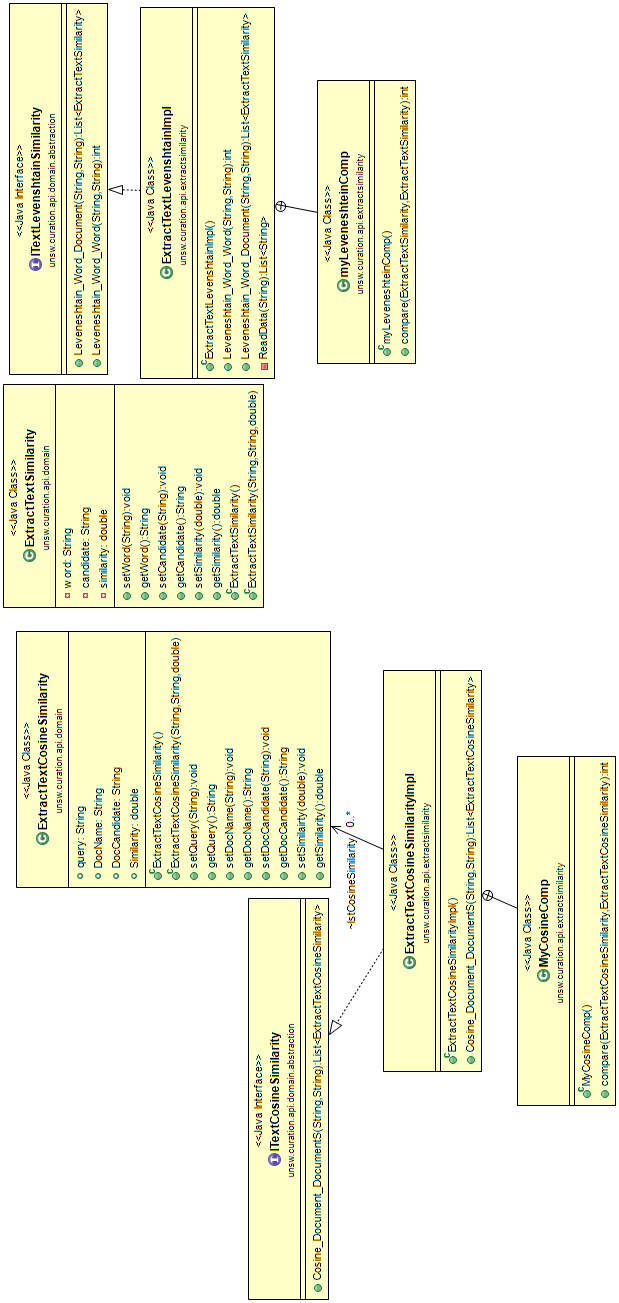}}
  \captionof{figure}{Class diagram of API}
  \label{fig:sim_class}
\end{center}

\subsection {API Implementation}
Similarity API consists of several libraries for computing the similarity of numerical and textual values. In the simplest form of processing the numerical values the API receives two vectors and returns a number as the measure of similarity. For more complex processing tasks, like computing the similarity of a set of vectors, the API returns a list, type of "ExtractNumberSimilarity" object. The list contains three properties; Vector1 (primary vector), Vector2 (extracted candidate vector) and score (similarity score between vector1 and vector2). For the textual contents the API has similar structure. For some similarity metrics, including soundex and  qgram the API receives two tokens (Word) and returns a number as the measure of similarity between tokens. In more complex processing tasks, like processing the similarity of files or sentences, it returns a list type of "ExtractTextSimilarity" object. The list contains three properties; word (primary token) candidate (extracted candidate token) and similarity (similarity score between the primary token and the candidate token). For processing the output of these methods, we have to create a list of \emph{ExtractNumberSimilarity} or \emph{ExtractTextSimilarity} and iterate the list using the \emph{"for"} loop statement.

\paragraph{Exception:}
For computing the cosine similarity of textual values, we need to create a list of objects type "ExtractTextCosineSimilarity" and for the tf-idf similarity metric we have to create a list of objects type of "ExtractTextTfIdfSimilarity". \emph{ExtractTextCosineSimilarty} library returns three properties for the retrieved data. \emph{DocName} (contains the original document), \emph{DocCandidate} (contains the extracted document) and similarity (returns the cosine similarity of 'DocName' and 'DocCandidate'). Also, tfidf similarity, returns a list of objects type of \emph{ExtractTextTfIdfSimilarity}, with three properties; sentence (query), \emph{SimilarSentence} (contains the candidate sentence extracted for the the given query) and score (returns the similarity of 'sentence' and the 'similarSentence').
\paragraph{Notice:}
For computing the similarity of numerical vectors, we used a vector file. Creating the vector file is simple, just needs to add the vectors in a separate line. Figure ~\ref{fig:sim_vec_file} shows a sample vector file.

\begin{center}

 \begin{tabular}{c}
  \setlength\fboxsep{5pt}
  \setlength\fboxrule{0pt}
  \fbox{\includegraphics{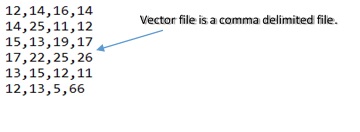}}
\end{tabular}
\captionof{figure}{Structure of vector file}
\label{fig:sim_vec_file}
\end{center}

\paragraph{Notice:}
Numerical libraries contains three methods. In the simplest form a method for computing the similarity of two vectors, a method for computing the similarity of a set of vectors together and eventually, a method for computing the similarity of a vector with a collection of vectors in a vector file. choosing the proper method is straightforward. The name of each method consist of three parts \textless Similarity Metric, Vector, Vector VectorS\textgreater. For example, for computing the cosine similarity of two vectors call \textless cosine\textunderscore vector\textunderscore vector \textgreater method or for computing the dice similarity a vector with a collection of vectors call \textless Dice\textunderscore vector\textunderscore vectorS \textgreater method. Table ~\ref{tbl:sim_lib} shows the input parameters and output parameters of numerical similarity methods.

\begin{center}

 \begin{tabular}{c}
  \setlength\fboxsep{0pt}
  \setlength\fboxrule{0.0pt}
  \fbox{\includegraphics[scale=0.45]{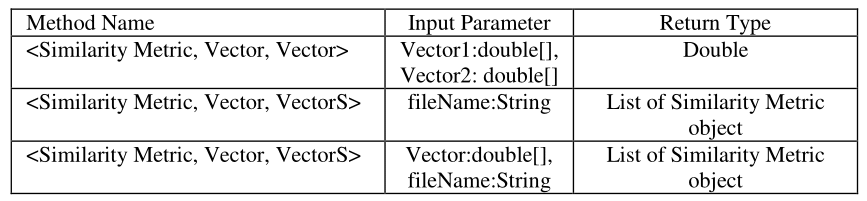}}
\end{tabular}
\captionof{table}{Structure of methods in the libraries }
\label{tbl:sim_lib}
\end{center}

Tables ~\ref{tbl:sim_num} - ~\ref{tbl:sim_num2} shows the structure of libraries provided for computing the similarity of numerical vectors. "DiceCoefficient" library computes dice similarity of input vectors, "NumberCosine" and Euclidean libraries compute cosine and euclidean similarity and JaccardCoefficient library computes jaccard similarity of vectors. Following shows the structure of libraries and their methods.

\begin{center}

 \begin{tabular}{c}
  \setlength\fboxsep{0pt}
  \setlength\fboxrule{0.0pt}
  \fbox{\includegraphics[scale=0.45]{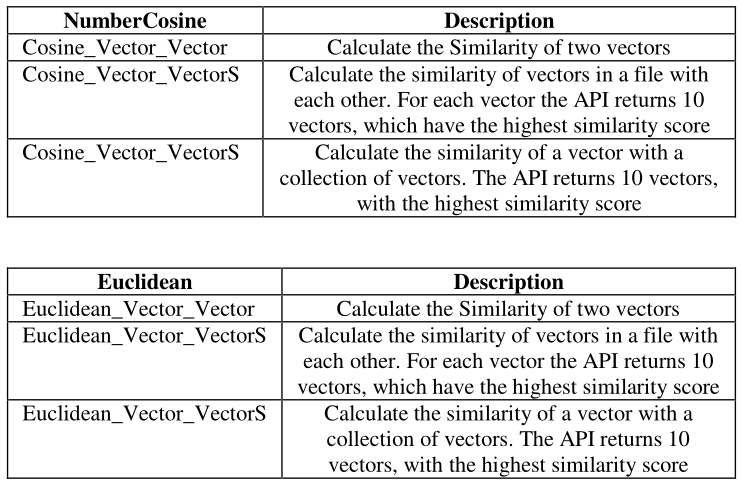}}
\end{tabular}
\captionof{table}{Detailed structure of numeric libraries}
\label{tbl:sim_num}
\end{center}
\
\begin{center}

 \begin{tabular}{c}
  \setlength\fboxsep{0pt}
  \setlength\fboxrule{0.0pt}
  \fbox{\includegraphics[scale=0.45]{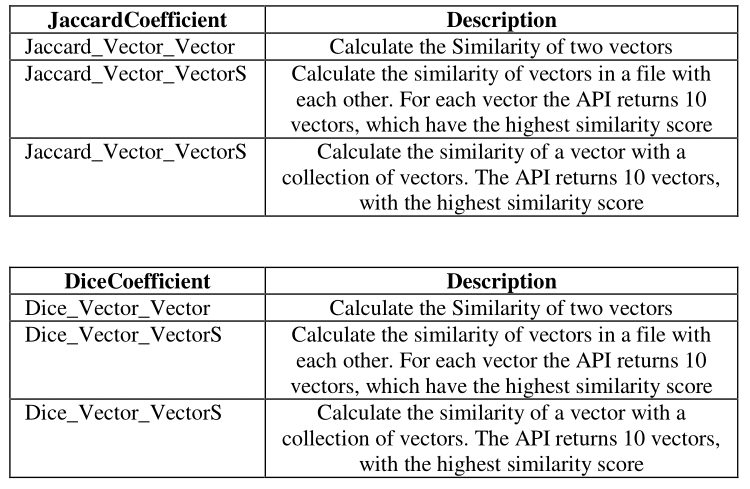}}
\end{tabular}
\captionof{table}{Detailed structure of numeric libraries}
\label{tbl:sim_num2}
\end{center}

\subsubsection{Similarity between String Values}
The similarity API contains several libraries for calculating the similarity of textual values. The API, computes Cosine, Jaccard, TF-IDF, Dice and Leveneshtein, soundex, and qgram similarity of textual values. Cosine and TFIDF similarity metrics are implemented based on the LUCENE information retrieval system \footnote{\path{https://lucene.apache.org}}. Levenshtein  uses dynamic programming technique and returns the number of edit distances require between words.
\paragraph{Notice:}
For computing the cosine similarity of documents the library contains a method named Cosine\textunderscore Document\textunderscore DocumentS, which receives two parameters. A file name (query) and a directory path (a set of candidate files). Both query file and directory candidate path must be in the same location. Tables ~\ref{tbl:sim_text} - ~\ref{tbl:sim_text2} shows the structure of libraries.

\begin{center}

 \begin{tabular}{c}
  \setlength\fboxsep{0pt}
  \setlength\fboxrule{0.0pt}
  \fbox{\includegraphics[scale=0.45]{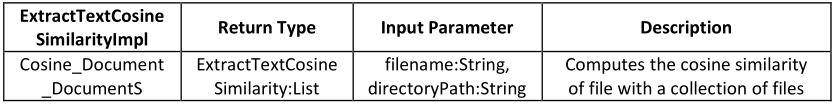}}
\end{tabular}
\captionof{table}{Structure of cosine libraries}
\label{tbl:sim_text}
\end{center}

\begin{center}

 \begin{tabular}{c}
  \setlength\fboxsep{0pt}
  \setlength\fboxrule{0.0pt}
  \fbox{\includegraphics[scale=0.45]{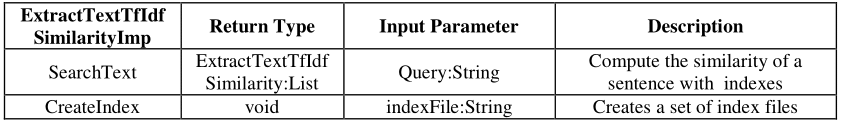}}
\end{tabular}
\captionof{table}{Structure of tfidf libraries}
\end{center}

\begin{center}

 \begin{tabular}{c}
  \setlength\fboxsep{0pt}
  \setlength\fboxrule{0.0pt}
  \fbox{\includegraphics[scale=0.45]{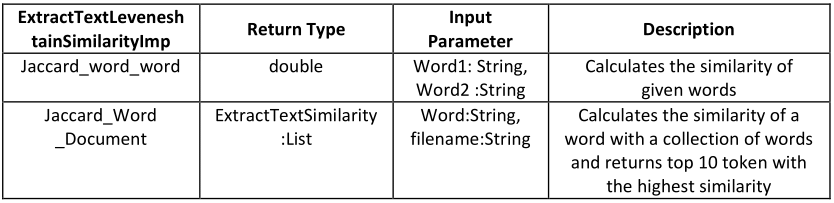}}
\end{tabular}
\captionof{table}{Structure of leveneshtain libraries}
\end{center}

\begin{center}

 \begin{tabular}{c}
  \setlength\fboxsep{0pt}
  \setlength\fboxrule{0.0pt}
  \fbox{\includegraphics[scale=0.45]{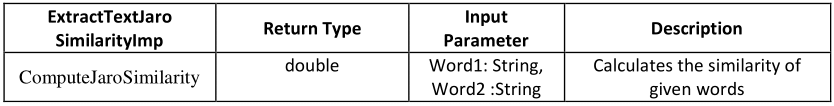}}
\end{tabular}
\captionof{table}{Structure of jaro libraries}
\end{center}

\begin{center}

 \begin{tabular}{c}
  \setlength\fboxsep{0pt}
  \setlength\fboxrule{0.0pt}
  \fbox{\includegraphics[scale=0.45]{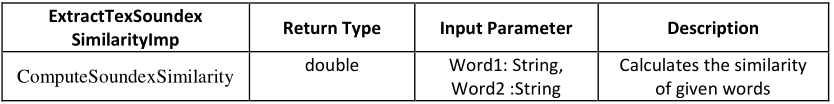}}
\end{tabular}
\captionof{table}{Structure of soundex libraries}
\end{center}

\begin{center}

 \begin{tabular}{c}
  \setlength\fboxsep{0pt}
  \setlength\fboxrule{0.0pt}
  \fbox{\includegraphics[scale=0.45]{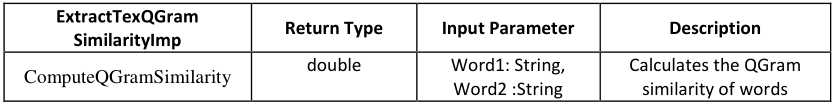}}
\end{tabular}
\captionof{table}{Structure of q-gram libraries}
\end{center}

\begin{center}

 \begin{tabular}{c}
  \setlength\fboxsep{0pt}
  \setlength\fboxrule{0.0pt}
  \fbox{\includegraphics[scale=0.45]{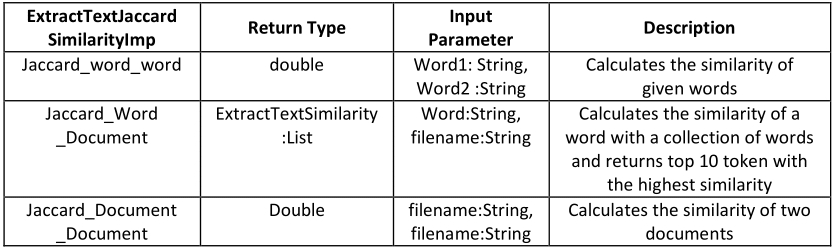}}
\end{tabular}
\captionof{table}{Structure of jaccard libraries}
\label{tbl:sim_text2}
\end{center}

\subsection{Evaluation}
In this section we present a part of output that obtained from Similarity API. In table ~\ref{tbl:sim_output} we have presented the output of computing the cosine and euclidean distance between different types of data. The first row of table shows the output of computing the cosine similarity of two vectors. The second row shows the output of computing the cosine similarity of a vector with a collection of vectors. The third row of table shows the output of computing the similarity of a set of vectors together, and the last row of table shows the output of computing the Euclidean distance among two vectors. The sample code for using the similarity API is available in more detail in the source code of project (unsw.curation.api.run package). Also in order to use the API refer to section 9.5 (User Guide).

\begin{center}

 \begin{tabular}{c}
  \setlength\fboxsep{0pt}
  \setlength\fboxrule{0.0pt}
  \fbox{\includegraphics[scale=0.45]{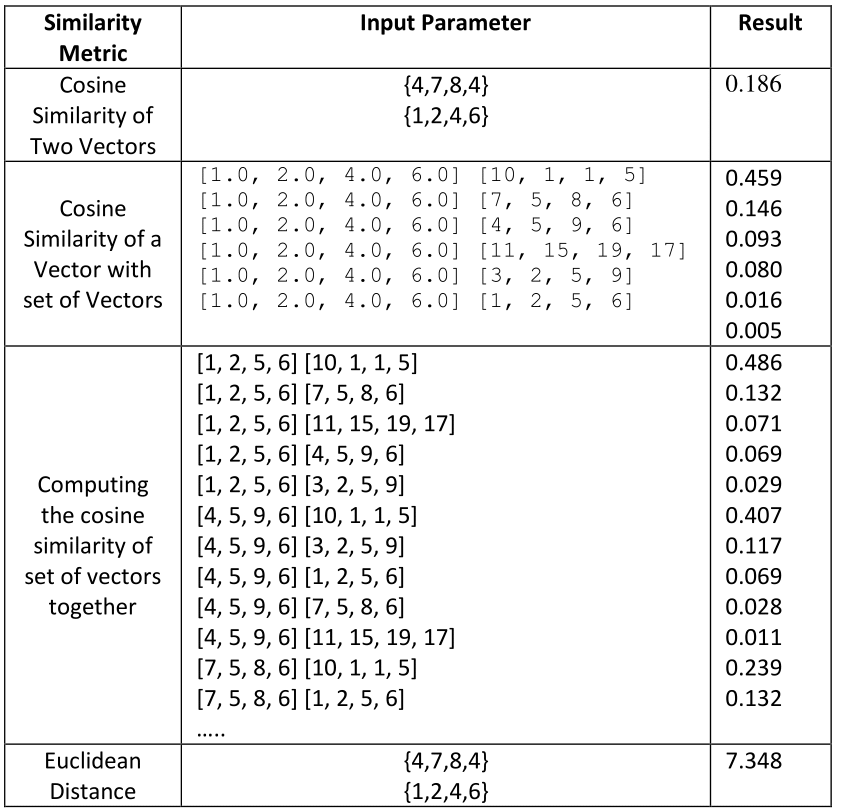}}
\end{tabular}
\captionof{table}{The output obtained from Similarity API}
\label{tbl:sim_output}
\end{center}

\subsection{User Guide}
This section shows the list of libraries are available in the API. For computing the similarity of numerical vectors the API contains the following libraries; Euclidean (ExtractNumberEuclideanSimilarity), cosine (ExtractNumberCosineSimilarity), Dice (ExtractNumberDiceSimilarity) and Jaccard (ExtractNumberJaccardSimilarity). Also the API contains the following libraries for computing the similarity of textual values. Leveneshtain (ExtractTextLevenshtainSimilarity), Cosine (ExtractTextCosineSimilarity), TF-IDF (ExtractTextTfIdfSimilarity), QGram (ExtractTextQGramSimilarity), Jaro (ExtractTextJaroSimilarity), Soundex (ExtractTextSoundexSimilarity) and Jaccard (ExtractTextJaccardSimilarity).

Using the Similarity API libraries are simple, just follow the following steps.
\begin{enumerate}
\item Open the project in Eclipse IDE (File -\textgreater Import -\textgreater Maven -\textgreater Existing Maven Projects -\textgreater Next -\textgreater click browse and select the folder that is the root of the maven project -\textgreater click Ok).
\item Create a new class and add the following method to your class -\textgreater public static void main(String [] args).
\item  Import unsw.curation.api.extractsimilarity.
\item  Create an instance of one the similarity libraries.
\item  Call one of the provided methods to compute the similarity.
\item  Iterate the results using the "for" loop statement.
\end{enumerate}
\textbf{Example:}\newline
Computing the cosine similarity of two vectors.

\begin{center}
 \begin{tabular}{c}
  \setlength\fboxsep{5pt}
  \setlength\fboxrule{0pt}
  \fbox{\includegraphics[scale=0.45]{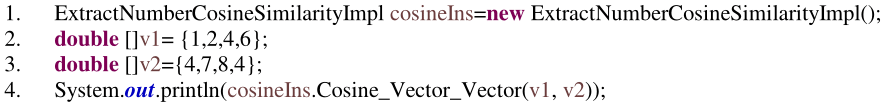}}
\end{tabular}
\end{center}
\textbf{Example:}\newline
Computing the cosine similarity of a vector with a collection of vectors.

\begin{center}
 \begin{tabular}{c}
  \setlength\fboxsep{5pt}
  \setlength\fboxrule{0pt}
  \fbox{\includegraphics[scale=0.45]{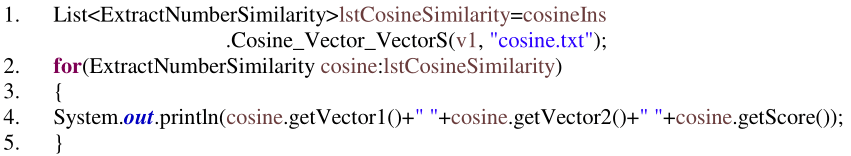}}
\end{tabular}
\end{center}
\textbf{Example:}\newline
Computing the cosine similarity of a collection of vectors in a vector file.
\begin{center}
 \begin{tabular}{c}
  \setlength\fboxsep{5pt}
  \setlength\fboxrule{0pt}
  \fbox{\includegraphics[scale=0.45]{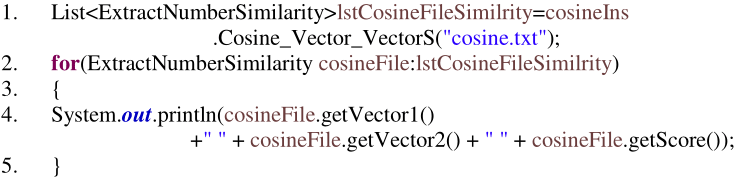}}
\end{tabular}
\end{center}
\textbf{Example:}\newline
Computing the Euclidean distance of two vectors (line 2) and line three computes the euclidean distance of a vector with a set of vectors.
\begin{center}
 \begin{tabular}{c}
  \setlength\fboxsep{5pt}
  \setlength\fboxrule{0pt}
  \fbox{\includegraphics[scale=0.45]{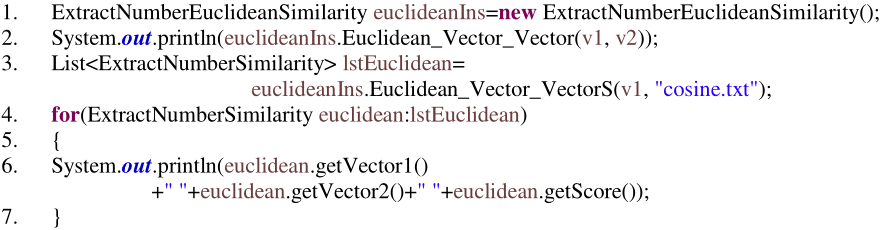}}
\end{tabular}
\end{center}

\begin{center}
 \begin{tabular}{c}
  \setlength\fboxsep{5pt}
  \setlength\fboxrule{0pt}
  \fbox{\includegraphics[scale=0.45]{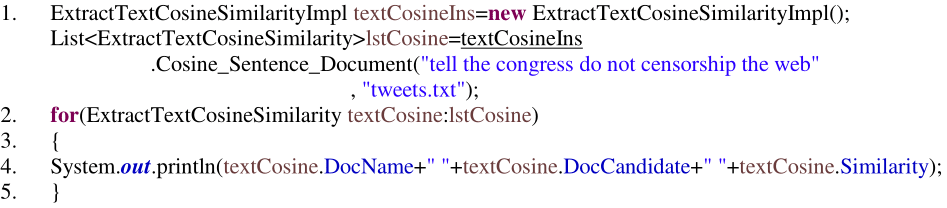}}
\end{tabular}
\end{center}
\textbf{Example:}\newline
Compute the tf-idf similarity of a query with a collection of sentences.
\begin{center}
 \begin{tabular}{c}
  \setlength\fboxsep{5pt}
  \setlength\fboxrule{0pt}
  \fbox{\includegraphics[scale=0.45]{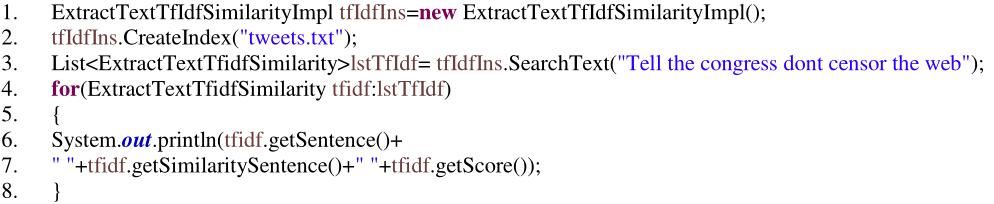}}
\end{tabular}
\end{center}
\paragraph{Notice:} For computing the tf-idf similarity, we need to create a set of index files. Therefore, first call the "CreateIndex" method, If the index files have not created yet. Then call the "searchText" method.
\paragraph{Notice:} every time we call the "CreateIndex" method, the API deletes the previous index files and creates a new set of indexes. For preserving the previous index files,  we don't need to call the "CreateIndex" method, every time.\newline\newline
Computes the cosine similarity of a file with collection of files.
\begin{center}
 \begin{tabular}{c}
  \setlength\fboxsep{5pt}
  \setlength\fboxrule{0pt}
  \fbox{\includegraphics[scale=0.45]{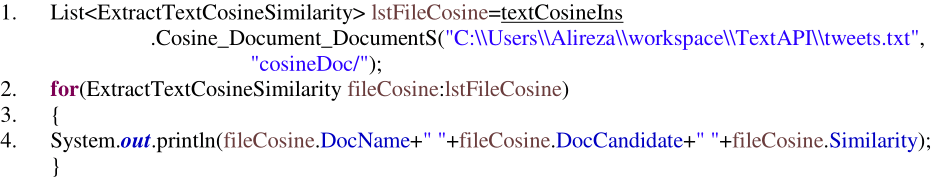}}
\end{tabular}
\end{center}

\newpage

\section{Appendix: STEM API}

\subsection{Introduction}
A stem is a form to which affixes can be attached~\cite{sampson2005language}. For example, the word friendships contains the stem friend, to which the derivational suffix -ship is attached to form a new stem friendship, to which the inflectional suffix -s is attached. To assist analysts understand and analyze the textual context, it will be important to extract derived form of the words in the text. Figure ~\ref{fig:stem_API} shows the structure of the API.
\begin{center}

 \begin{tabular}{c}
  \setlength\fboxsep{0pt}
  \setlength\fboxrule{0.0pt}
  \fbox{\includegraphics[scale=0.45]{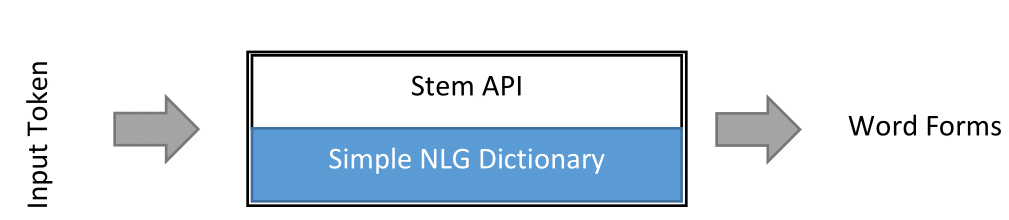}}
\end{tabular}
\captionof{figure}{Structure of Stem API}
\label{fig:stem_API}
\end{center}

\subsection{API Specification}
Stem is an API for extracting the derived forms of words (Adjective, Adverb, Noun and Verb). The API returns different forms of a token and implemented based on an API, named Simple NLG\footnote{\path{https://github.com/simplenlg/simplenlg}}. Figure ~\ref{fig:stem_class} shows the class diagram of the API.
\begin{center}

 \begin{tabular}{c}
  \setlength\fboxsep{0pt}
  \setlength\fboxrule{0.0pt}
  \fbox{\includegraphics[scale=0.8]{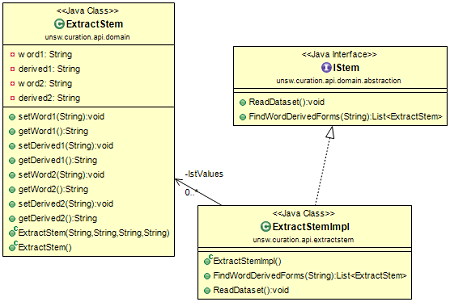}}
\end{tabular}
\captionof{figure}{Class diagram of Stem API}
\label{fig:stem_class}
\end{center}
\subsection{API Implementation}
Using the API is straightforward, at the core of the API is a library named ExtractStemImpl. The library, contains several methods for extracting the derived forms of words. "FindWordDerivedForms" method receives a word and returns the Adjectives, Adverb, Noun and Verbs (if exists). "FindSentenceDerivedForms" method receives a sentence and return the word forms of keywords in the sentence. Also, "FindFileDerivedForms" method extracts the word forms of keywords exist in a file. Table ~\ref{tbl:stem_lib} shows the structure of API.

\begin{center}

 \begin{tabular}{||c c||}
 \hline
 Method Name &  Description\\
  \hline\hline
  FindWordDerivedForm& return the derived form of input word\\
  \hline
\end{tabular}
\captionof{table}{Structure of Stem API library}
\label{tbl:stem_lib}
\end{center}

\paragraph{Notice:}
The API needs a database of word forms (Adjective, Adverb, Noun, Verbs). The database located at the root of project. \\
\subsection{Evaluation}
 Stem API is a simple but useful API for extracting the "wordforms". The API receive a token and returns all other forms of token. Table ~\ref{tbl:stem_output} shows the word "PLAY" is added to API as the input parameter and the API returned the other form of token.
\begin{center}

 \begin{tabular}{c}
  \setlength\fboxsep{0pt}
  \setlength\fboxrule{0.0pt}
  \fbox{\includegraphics[scale=0.45]{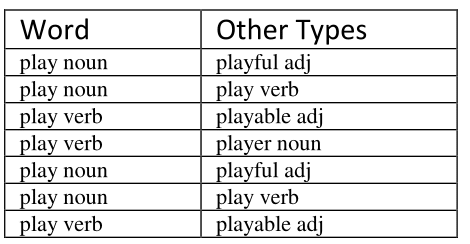}}
\end{tabular}
\captionof{table}{The output obtained from stem API}
\label{tbl:stem_output}
\end{center}
\subsection{Experiments}
In order to use the Stem API follow the following steps, where a sample code for extracting the derived form of an input word has been illustrated.
\begin{enumerate}
\item Import 'unsw.curation.api.extractstem.ExtractStemImpl library to your project.
\item Create an instance of ExtractStemImp library
\item Call ReadDataset method (For loading the database file)
\item Call FindWordDerivedForm method for extracting the word forms.
\end{enumerate}
\textbf{Example:}\newline
\begin{center}
 \begin{tabular}{c}
  \setlength\fboxsep{5pt}
  \setlength\fboxrule{0pt}
  \fbox{\includegraphics[scale=0.38]{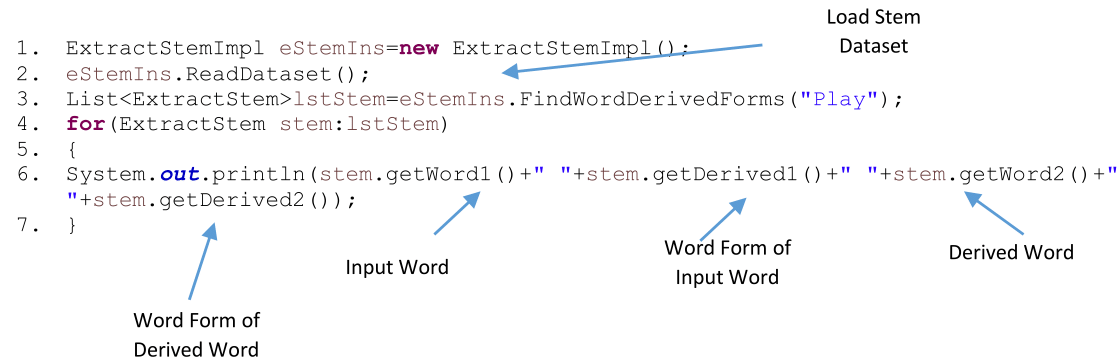}}
\end{tabular}
\end{center}

\newpage

\section{Appendix: Classification API}

\subsection{Introduction}
Data mining is widely used in variety of techniques including medicine, finance and weather forecasting. Data mining goal is the extraction of a set of raw data and transforming them into the useful knowledge. In this section we introduce an API for text classification. Classification is a sub task of data mining that assigns a label to each item in a collection. Classification tries to accurately predict the target label for a collection of unseen data. A classification task begins with data in, which the class label are known. For example, a classification model for predicting the weather could be developed based on data collected from previous days during a period of time.
The simplest type of classification problem is binary classification. In binary classification, the target attribute has only two possible values: for example, sunny or rainy. Multiclass targets have more than two values: for example, sunny, rainy, cloudy.
The model creation in datamining consists of two steps; training and testing. In the training phase, a classification algorithm finds relationships between the values of the predictors and the values of the target. Different classification algorithms use different techniques for finding relationships. Then the classification models are tested by comparing the predicted values to known target values in a set of test data. The historical data for a classification project is typically divided into two data sets; one for building the model; the other for testing the model. The structure of API is presented in figure ~\ref{fig:clas_ext}.

\begin{center}

 \begin{tabular}{c}
  \setlength\fboxsep{5pt}
  \setlength\fboxrule{0pt}
  \fbox{\includegraphics[scale=0.9]{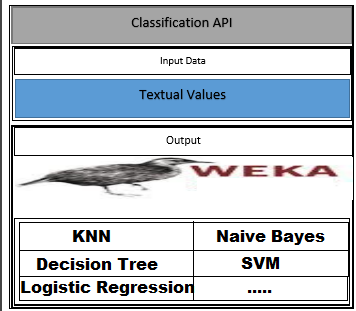}}
\end{tabular}
\captionof{figure}{Structure of Classification API.}
\label{fig:clas_ext}
\end{center}

\subsection{API Specification}
Classification API is a Java based API and contains several algorithms for classification task. The API is created on top of WEKA library and contains Naive Bayes, Decision Tree, Logistic Regression, SVM, Random Forest, KNN and Neural Network algorithms. The class diagram is presented in figures ~\ref{fig:clas_clas}.

\subsection{API Implementation}
The API contains a set of algorithms for classifying text. "TextClassifierImpl" and "EvaluateClassifier" are the libraries, placed at the core of API. \emph{TextClassifierImpl} library classifies the text and \emph{EvaluateClassifier} library contains a set of methods for computing precision, recall and accuracy of models. Table ~\ref{tbl:clas_method} shows the the structure of methods provided for classifying text and table ~\ref{tbl:clas_eval} shows the the methods implemented for evaluating model performance.

\begin{center}
  \setlength\fboxsep{5pt}
  \setlength\fboxrule{0pt}
  \fbox{\includegraphics[scale=0.55]{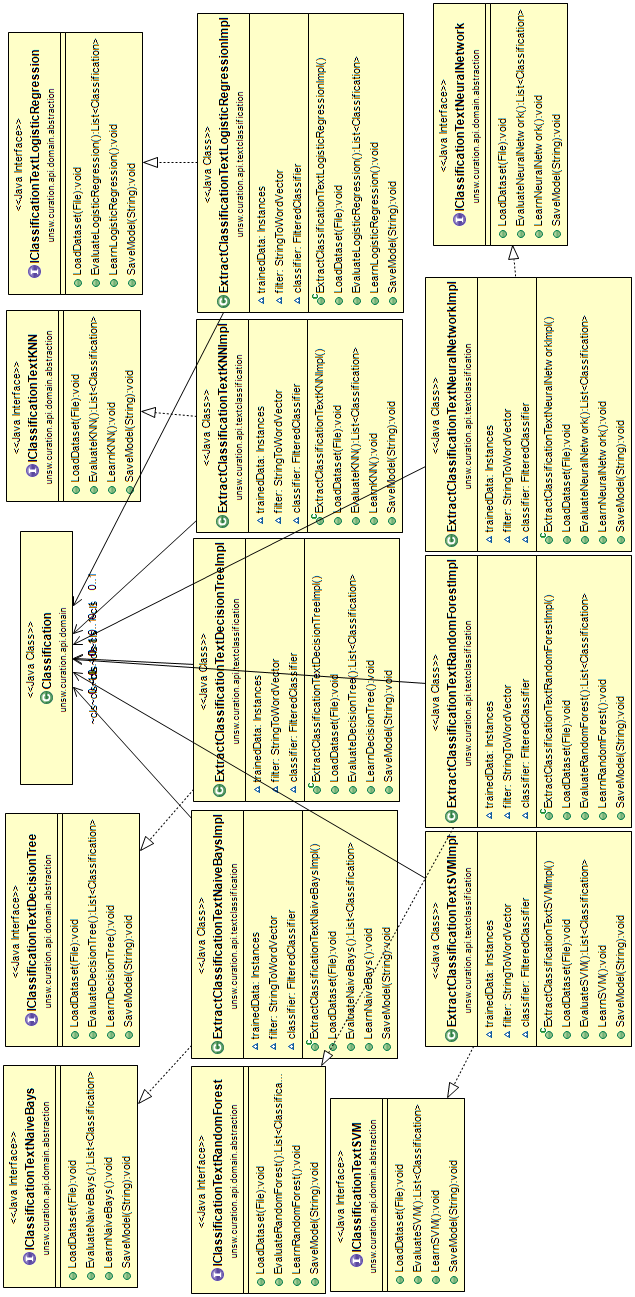}}
  \captionof{figure}{Class Diagram of Classification API}
  \label{fig:clas_clas}
\end{center}

\begin{center}
 \begin{tabular}{c}
  \setlength\fboxsep{5pt}
  \setlength\fboxrule{0pt}
  \fbox{\includegraphics[scale=0.45]{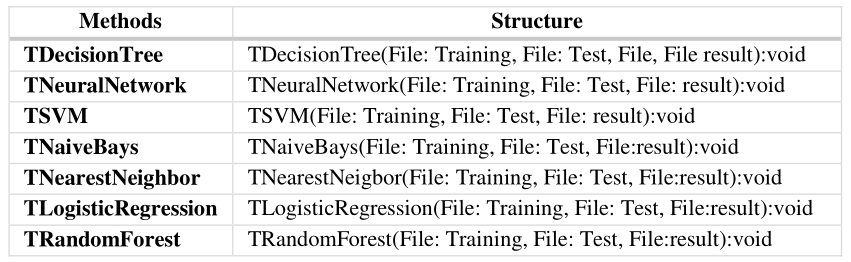}}
\end{tabular}
\captionof{table}{methods provided for Classification}
\label{tbl:clas_method}
\end{center}

\begin{center}

 \begin{tabular}{c}
  \setlength\fboxsep{5pt}
  \setlength\fboxrule{0pt}
  \fbox{\includegraphics[scale=0.45]{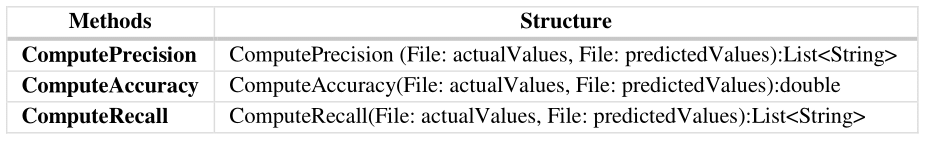}}
\end{tabular}
\captionof{table}{methods provided for model evaluation}
\label{tbl:clas_eval}
\end{center}

As is presented in table ~\ref{tbl:clas_method} the API contains seven algorithms, each method receives three parameters. The first parameter, indicate the \emph{training file path}, the second one indicates the \emph{test file path} and the last one shows the path of \emph{output result}. Also, table ~\ref{tbl:clas_eval} shows the structure of "EvaluateClassifier" library, which contains "ComputePrecision", "ComputeRecall" and "ComputeAccuracy" methods. Every methods receives two parameters, the first parameter is a file with the predicted labels and the second one is the path of actual labels.
\subsubsection{ Training and test file format}
The API receives two .ARFF files for training and testing.  An "ARFF" files contains two sections, header and data. Header contains the name of relation, a list of attributes (the columns in the data), and their types. Data contains the values of each attribute. The @RELATION, @ATTRIBUTE and @DATA shows the valid declaration of an "ARFF" file. Following shows the steps of creating a test and a train ARFF file.

\begin{enumerate}
\item  Add @Relation with an optional name (@Relation Name).
\item  Add @attribute (Setting field name and attributes).
\item  Add @Data (add the training data (data must be comma delimited and the number of columns should be equal with the number of attributes)).
\end{enumerate}
\newpage
\subsubsection{Creating a train ARFF file}
Figure ~\ref{fig:clas_arff} shows the steps for creating an ARFF file.

\begin{center}
 \begin{tabular}{c}
  \setlength\fboxsep{5pt}
  \setlength\fboxrule{0pt}
  \fbox{\includegraphics{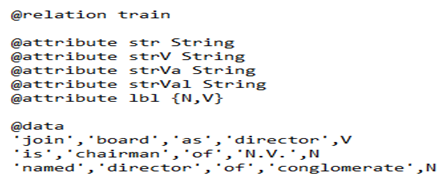}}
\end{tabular}
\captionof{figure}{sample training Arff file}
\label{fig:clas_arff}
\end{center}

\paragraph{Notice:} The \emph{last attribute} in the training file should be the \emph{class label}.
\subsubsection{Creating a test ARFF file}
Figure ~\ref{fig:clas_test_arff} shows the steps needs for creating a test ARFF file.

\begin{center}
 \begin{tabular}{c}
  \setlength\fboxsep{5pt}
  \setlength\fboxrule{0pt}
  \fbox{\includegraphics{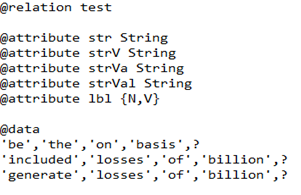}}
\end{tabular}
\captionof{figure}{sample test Arff file}
\label{fig:clas_test_arff}
\end{center}

\begin{enumerate}
\item  Add @Relation and a custom name to the first line of your file.
\item  Add @Attribute (@Attribute  FieldName Type).
\item  Add @Data (In this section add the test data (data must be comma delimited)).
\end{enumerate}

\paragraph{Notice:}
For the test file you shouldn't set the class label. Use question mark instead of class label
\paragraph{Notice:}In the test file the \emph{last attribute} should be the class label.
\subsection{Evaluation}
In this section we present the output of Classification algorithms. We divided our data into two parts, test and train. We classify the data using KNN and Naive bayes algorithms. Table ~\ref{tbl:clas_output} shows the output of classification API. Also for using the API refer to the section 11.5 (User Guide).
\begin{center}

 \begin{tabular}{c}
  \setlength\fboxsep{0pt}
  \setlength\fboxrule{0.0pt}
  \fbox{\includegraphics[scale=0.45]{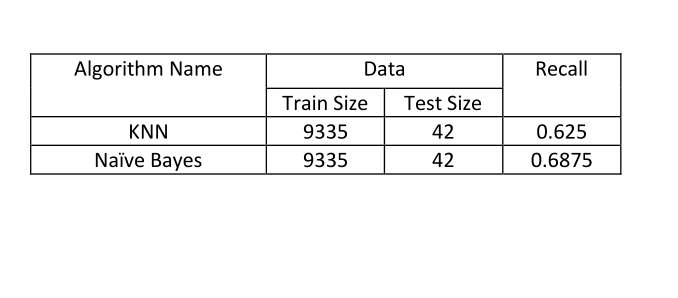}}
\end{tabular}
\captionof{table}{The output obtained from Classification API}
\label{tbl:clas_output}
\end{center}
\subsection{User Guide}
Using the classification API is straightforward, for predicting the labels of the test data, create an instance of "TextClassifierImpl" library and call one of the available methods. Also for evaluating the performance of the classifiers, create an instance of "EvaluateClassifier" library and call one of the provided methods. Following shows the steps need for creating a model and evaluating it's performance.

\begin{enumerate}
\item import unsw.curation.api.textclassification.TextClassifierImpl.
\item create an instance of TextClassifierImpl library.
\item call one of the methods provided as part of library (every methods receives three parameters (training file path, test file path and output result)).
\end{enumerate}
For evaluating the performance of created model follow the following steps:

\begin{enumerate}
\item Open the project in Eclipse IDE (File -\textgreater Import -\textgreater Maven -\textgreater Existing Maven Projects -\textgreater Next -\textgreater Click browse and select the folder that is the root of the Maven project -\textgreater Click Ok)
\item Create a new class and add the following method to your class -\textgreater public static void main(String [] args)
\item import unsw.curation.api.textclassification.EvaluateClassifier
\item create an instance of EvaluateClassifier library
\item call one of the methods provided as part of library (for evaluating the performance of a classifier, every method receive two parameters (actual test label file path, predicted labels).
\end{enumerate}

Following shows how to classify an input data using KNN and Naive Bayes algorithms. The process of classifying data using other algorithms are similar.\\
\textbf{Example:}\newline
\begin{center}
\begin{tabular}{c}
  \setlength\fboxsep{5pt}
  \setlength\fboxrule{0pt}
  \fbox{\includegraphics[scale=0.45]{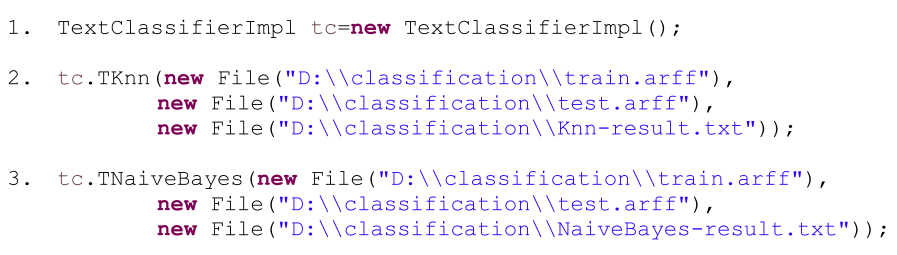}}
\end{tabular}
\end{center}
\textbf{Example:}\newline
\paragraph{Evaluating the performance of models}
\begin{center}
\begin{tabular}{c}
  \setlength\fboxsep{5pt}
  \setlength\fboxrule{0pt}
  \fbox{\includegraphics[scale=0.38]{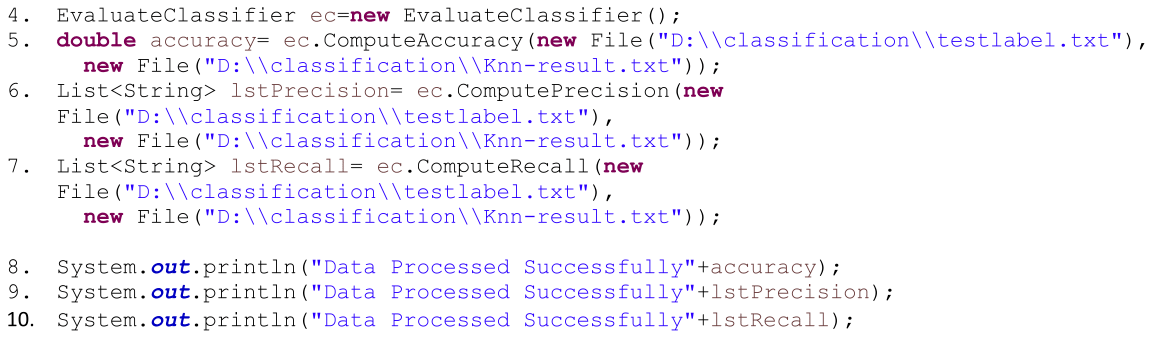}}
\end{tabular}
\end{center}

\paragraph{Notice:} For Evaluating the performance of a classifier the length of both actual test labels and predicted labels must be equal.
\newpage

\section{Appendix: Linking API}

\subsection{Introduction}
Entity Linking is the task of determining the identity of entities mentioned in a text. For example, given the sentence "Paris is the capital of France", the idea is to determine that "Paris" refers to the city of Paris and not to Paris Hilton or any other entity that could be referred as "Paris"\footnote{\path{www.wikipedia.com}}.
\subsection{API specificaton}
Linking API, is designed for extracting knowledge from "Wikidata", "Google Knowledge Graph" and "ConceptNet". Google Knowledge Graph is a knowledge base used by Google  for improving the google search results. Wikidata is a knowledge base operated by the Wikimedia Foundation. It is intended to provide a common source of data which can be used by Wikimedia projects such as Wikipedia.  ConceptNet is a semantic network containing the informations that computers need to know about the world. It is built from nodes(Concept) representing words or short phrases of natural language, and labeled relationships between them. These are the kinds of relationships computers need to know to search for information better, answer questions, and understand people's goals.
\subsection{API Implementation}
Linking API has simple structure, it contains three libraries named WikiData, GoogleKnowledgeGraph and ConceptNet. The first two libraries are responsible for extracting knowledge from google knowledge graph and wikidata and the third one is designed for extracting knowledge from ConceptNet. Both WikiData and GoogleKnowledgeGraph  libraries receive a list of tokens and return a JSON string that contains the information about the given entity. Also, ConceptNet library contains a set of methods for extracting the relation and associations between objects. Table ~\ref{tbl:link_struct} shows the structure of API.
\subsubsection{Wikidata and Google Knowledge Graph}
Using WikiData and GoogleKnowledgeGraph libraries are simple. WikiData library contains a method named "ParseWikidata",  which receives an entity and search the WikiData for the existence of an entity. Also "GoogleKnowledgeGraph" library contains a method named "ParseGoogleKnowledgeGraph" for linking an entity with an object in Google knowledge Graph.
\subsubsection{ConceptNet}
ConceptNet is an other library, for extracting the relations and associations among the words. following shows the structure of methods available in the library.

\begin{enumerate}
\item  ConceptNetLookUp:  returns the list of edges that include the object.
\item  ConceptNetPartOfSearch:  allows searching ConceptNet edges for multiple requirements.
\item  ConceptNetAssociation: returns the most similar concepts to a given concept.
\item  ConceptNetAssociationWords: returns the most similar concepts to a given concept.
\end{enumerate}

\begin{center}

 \begin{tabular}{||c c||}
 \hline
 Method Name &  Input Type\\
  \hline\hline
  ConceptNetLookUp & String:Token, Int:Count\\
  \hline
ConceptNetPartOfSearch & String:Token\\
\hline
ConceptNetAssociation & String:Token\\
\hline
ConceptNetAssociationWords & Array:Tokens\\
\hline
\end{tabular}
\captionof{table}{Structure of Linking API}
\label{tbl:link_struct}
\end{center}
\subsection{Evaluation}
In this section we present different types of output that can be extract from linking API. In first two rows of table ~\ref{tbl:link_output} we presented the output of ConceptNet library for two tokens (the first one extract the words are related to shirt and the second one demonstrate the \emph{"PART OF"} relation and returns the tokens as a part of car. The other lines show the outputs from "Google Knowledge Graph" and WikiData libraries. In here we just presented a part of output, you can refer to the source code of project to fully check the API.
\begin{center}
 \begin{tabular}{c}
  \setlength\fboxsep{0pt}
  \setlength\fboxrule{0.0pt}
  \fbox{\includegraphics[scale=0.45]{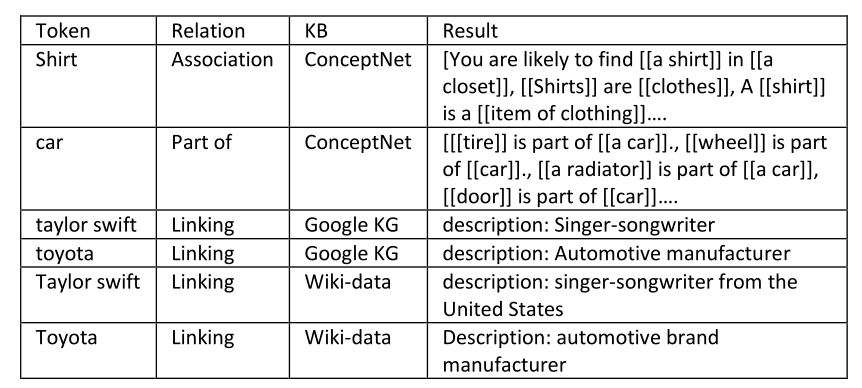}}
\end{tabular}
\captionof{table}{The output obtained from Linking API}
\label{tbl:link_output}
\end{center}
\subsection{User Guide}
In order to link an entity with Google Knowledge Graph or Wikidata just create an instance of their libraries and call "ParseWiki" or "ParseGoogleKnowledge" methods with the appropriate input values.\\
\textbf{Example:}\newline
Linking an entity with google knowledge graph and wikidata
\begin{center}
\begin{tabular}{c}
  \setlength\fboxsep{5pt}
  \setlength\fboxrule{0pt}
  \fbox{\includegraphics[scale=0.45]{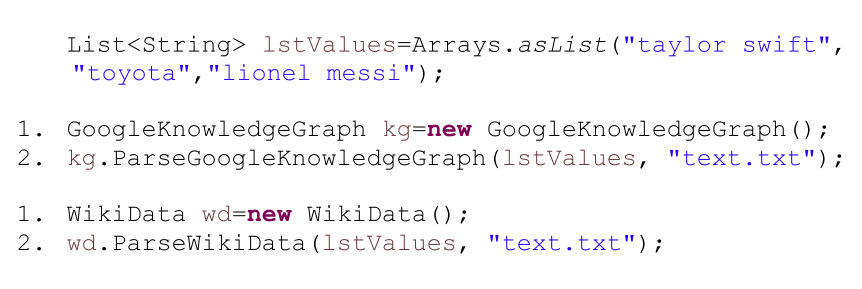}}
\end{tabular}
\end{center}
 Also,  ConceptNet library, contains several methods for extracting the relations between tokens.\\
 \textbf{Example:}\newline
\begin{center}

\begin{tabular}{c}
  \setlength\fboxsep{5pt}
  \setlength\fboxrule{0pt}
  \fbox{\includegraphics[scale=0.45]{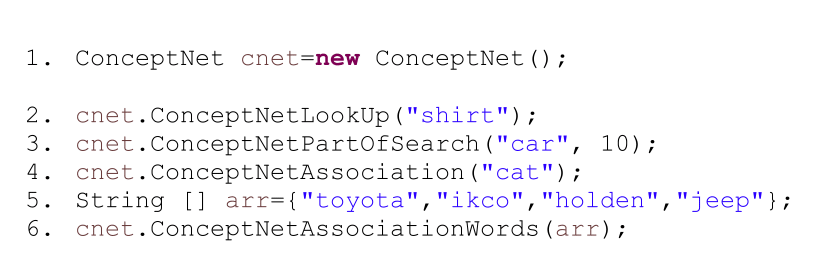}}
\end{tabular}
\end{center}
\begin{enumerate}
\item Line2 shows how to extract a list of edges that include the given object\\
\item Line3 shows how to return a list of objects with the (Part of relationship) with the given object\\
\item Line4 shows how to return a list of similar concept to the given concept\\
\item Line5 receives an array of objects and return a list of concepts similar to the given array.\\
\end{enumerate}
\newpage

\section{Appendix: Indexing API}

\subsection{Introduction}
Lucene is a full-text search library in Java which makes it easy to add search functionality to an application or website.
It does so by adding content to a full-text index and allows to perform queries on this index, returning results ranked by the relevance to the query.\newline Lucene is able to achieve fast search responses because, instead of searching the text directly, it searches an index instead. This would be the equivalent of retrieving pages in a book related to a keyword by searching the index at the back of a book, as opposed to searching the words in each page of the book.\footnote{\path{http://www.lucenetutorial.com/basic-concepts.html}}
\subsection{API Specification}
Indexing API performs high performance and full-featured text search and is suitable for applications that requires full-text search. The API receives a token or a phrase and returns the sentences that contains the given token. Also the user can set a slop number. Slop number determines the maximum distance that each part of a phrase can have in a sentence.
Example: If a user enter "mental health" as the input token, the API returns all the sentences contains mental and health, in a user defined window.

\subsection{API Implementation}
The API contains three classes, one for indexing, one for querying the index files and a class for returning the sentences contains the query. In order to search/index a piece of data we need to create a set of index files, then we have to search and retrieve the sentences that contains the query. For Using the Indexing API, follow the following steps.

\begin{enumerate}
\item Open the project in Eclipse IDE (File -\textgreater Import -\textgreater Maven -\textgreater Existing Maven Projects -\textgreater Next -\textgreater Click browse and select the folder that is the root of the Maven project -\textgreater Click Ok).
\item Create a new class and add the following method to your class -\textgreater public static void main(String [] args).
\item import unsw.curation.api.index.
\item Create an instance of Index library (Set the path of your data (data that will be indexed)).
\item Create a instance of SchIndData library.
\item call the IndexDocuments method for indexing the input values.
\item call the search method and specify the query for searching in the index file.
\end{enumerate}
\subsection{Evaluation}
Indexing API is a useful API for conducting the fast search. As presented in table ~\ref{tbl:index}, The word "mental health" has been added as the input token and the API returns the sentences that contains both mental and health. The source code and the User Guide provided in section 13.5 describes the steps of using the API.
\begin{center}

 \begin{tabular}{c}
  \setlength\fboxsep{0pt}
  \setlength\fboxrule{0.0pt}
  \fbox{\includegraphics[scale=0.45]{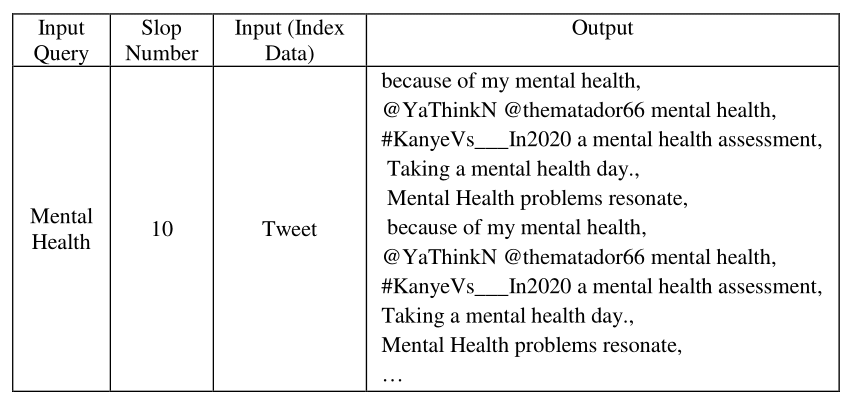}}
\end{tabular}
\captionof{table}{The output obtained from Indexing API}
\label{tbl:index}
\end{center}
\subsection{User Guide}
Using Indexing API is straightforward. Following shows a sample code for indexing and searching a text file.\\
\textbf{Example:}\newline

\begin{center}
\begin{tabular}{c}
  \setlength\fboxsep{5pt}
  \setlength\fboxrule{0pt}
  \fbox{\includegraphics[scale=0.45]{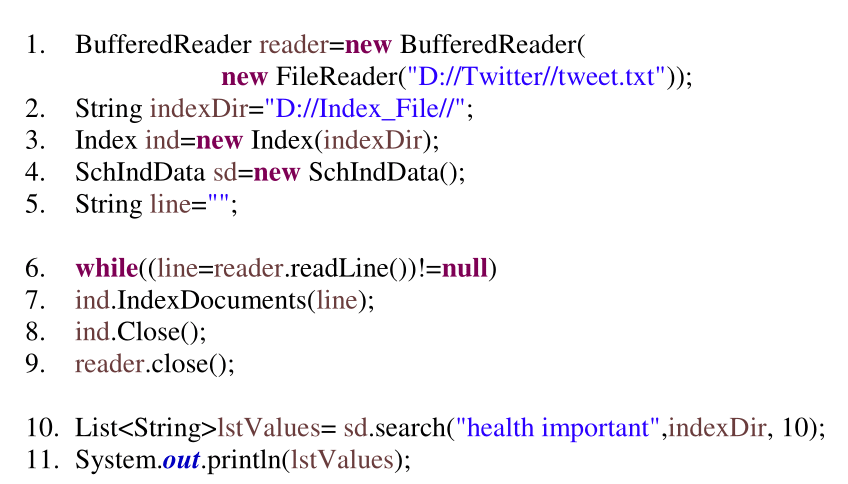}}
\end{tabular}
\end{center}
\paragraph{Notice:}
Line 1 reads a file named tweet.txt. Line 3 set the path of index directory. Line 4 creates an instance of SchIndData library and lines 6,7,8,9 are a loop statement for indexing the content of tweet.txt file and creating the index files. In line 10 the search method is called, which returns the sentences contains the search term. The first parameter in the search method is the search term (query), the second parameter is the index file directory and the third one is the slop distance.
\paragraph{Notice:} Indexing the data is an expensive task. Therefore, create the index file, whenever it is necessary. For using the previous index data, just use the code in lines 2,4,9,10.
\newpage

\section{Appendix: Demo Application: Twitter Extraction API}
In this section we will demonstrate a twitter streaming application, which created based on the  API named Twitter4j and the curation API's introduced in previous sections. The application named "Twitter Extraction" and extracts variety of information from twitter, including named entities, POS tags, synonyms, keywords.
\subsubsection{Twitter4j}
Twitter4J \footnote{\path{http://twitter4j.org/en/}} is an unofficial Java library for the Twitter API. With Twitter4J, you can easily integrate your Java application with the Twitter service.
\subsubsection{Implementation of Twitter Extraction API:}
Using the "Twitter Extraction" is straightforward and we will demonstrate it's usage in next sections. However, its necessary to register the API and having "OAuth" \footnote{\path{https://apps.twitter.com/}} for sending secure authorized requests to the Twitter. After registering the API and getting OAuthConsumerKey, OAUTHConsumerSecret, OAUTHAccessToken, OAUTHAccessTokenSecret codes, the API starts streaming data of twitter. Table ~\ref{tbl:twitter_element} shows the elements, that Twitter Extraction API extracts for each tweet. The API creates a XML file in the root of project named result.xml which contains both the data extracted using the curation API's (Named entity, Keyword, Url, Synonym and Stem) and the data extracted by twitter4j API.
\begin{center}

\begin{tabular}{c}
  \setlength\fboxsep{5pt}
  \setlength\fboxrule{0pt}
  \fbox{\includegraphics[scale=0.45]{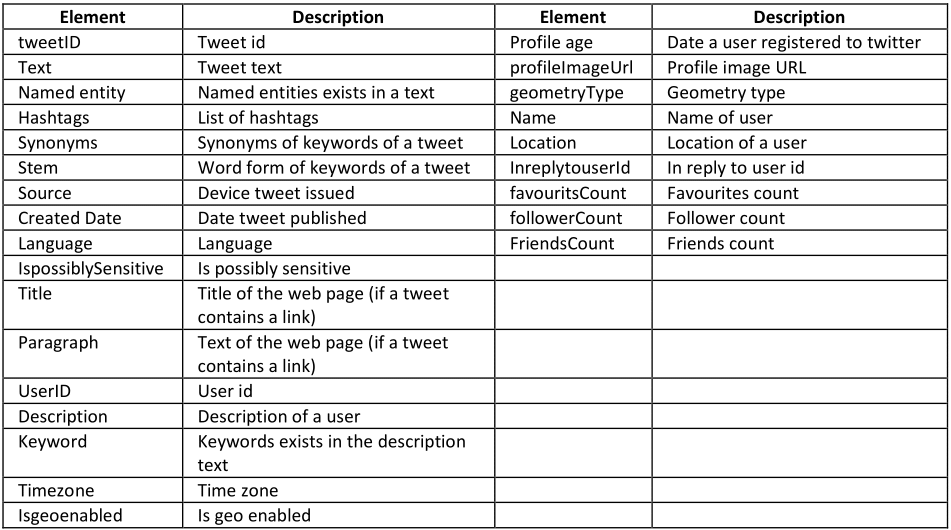}}
\end{tabular}
\captionof{table}{List of elements extracted for each tweets}
\label{tbl:twitter_element}
\end{center}
At the core of the API there is a library named TweetInfo, which enrich the tweets using the elements presented in table 14.1. The library contains two methods namely; "StartStreaming()" and "StartStreamingByKeyword()". The first one streams a user defined number of tweets and the second returns a specified number of tweets, which contain a user defined keyword.
\subsubsection{Using Twitter Extraction API:}
Using Twitter Extraction API is simple, just create an instance of a library named tweetInfo and call one of the methods provided as part of library.

\begin{enumerate}
\item  Import unsw.text.api.twitter package
\item  Create an Instance of library named TweetInfo
\item  Call themethods to compute the similarity
\end{enumerate}
\paragraph{Notice:}
The constructor of TweetInfo library, receives six parameters, table ~\ref{tbl:tweetInfo} describes the parameters more clearly. For creating an instance of TweetInfo library, fill the parameters with the necessary values.
\textbf{Example:}\newline
\begin{center}

\begin{tabular}{c}
  \setlength\fboxsep{5pt}
  \setlength\fboxrule{0pt}
  \fbox{\includegraphics[scale=0.45]{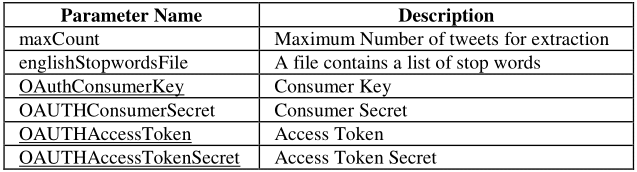}}
\end{tabular}
\captionof{table}{creating an instance of TwitterInfo library}
\label{tbl:tweetInfo}
\end{center}
\textbf{Example:}\newline
following figure shows a sample code for streaming 10 tweets.

\begin{center}
\begin{tabular}{c}
  \setlength\fboxsep{5pt}
  \setlength\fboxrule{0pt}
  \fbox{\includegraphics[scale=0.45]{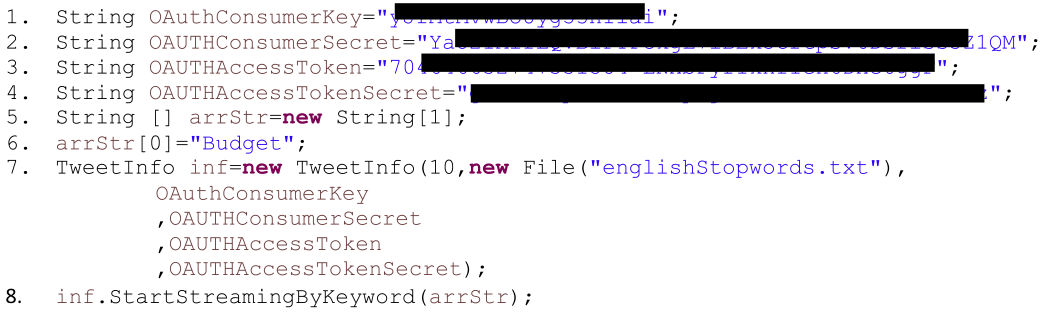}}
\end{tabular}
\end{center}
\newpage

\newpage

\noindent\textbf{Acknowledgements.}\\
We Acknowledge the Data to Decisions CRC (D2D CRC) and the Cooperative Research Centres Program for funding this research.

\newpage

%% bib
\bibliographystyle{plain}
\bibliography{sample}
\end{document}